\documentclass[12pt,preprint]{aastex}
\usepackage{lscape}

\slugcomment{to appear in the 1/1/2009 issue of the Astrophysical Journal}

\begin{document}{}

\def\t0{\theta_{\circ}}
\def\muo{\mu_{\circ}}
\def\sd{\partial}
\def\be{\begin{equation}}
\def\en{\end{equation}}
\def\bv{\bf v}
\def\bvo{\bf v_{\circ}}
\def\ro{r_{\circ}}
\def\rhoo{\rho_{\circ}}
\def\etal{et al.\ }
\def\msun{\,M_{\sun}}
\def\rsun{\,R_{\sun}}
\def\lsun{L_{\sun}}
\def\msunyr{M_{\sun} yr^{-1}}
\def\kms{\rm \, km \, s^{-1}}
\def\mdot{\dot{M}}
\def\Md{\dot{M}}
\def\curf{{\cal F}}
\def\ecs{erg cm^{-2} s^{-1}}
\def \haebe{HAeBe}
\def \mum {\,{\rm \mu m}}
\def \simali {{\sim\,}}
\def \K {\,{\rm K}}
\def \Angstrom     {\,{\rm \AA}}
\newcommand \g            {\,{\rm g}}
\newcommand \cm           {\,{\rm cm}}

\title{Silica in Protoplanetary Disks}

\author{B.A. Sargent\altaffilmark{1},
W.J. Forrest\altaffilmark{1},
C. Tayrien\altaffilmark{1},
M.K. McClure\altaffilmark{1},
A. Li\altaffilmark{2},
A.R. Basu\altaffilmark{3},
P. Manoj\altaffilmark{1},
D.M. Watson\altaffilmark{1},
C.J. Bohac\altaffilmark{1},
E. Furlan\altaffilmark{4},
K.H. Kim\altaffilmark{1},
J.D. Green\altaffilmark{1},
G.C. Sloan\altaffilmark{5}}

\altaffiltext{1}{Department of Physics and Astronomy, University of 
                 Rochester, Rochester, NY 14627;
                 {\sf bsargent@astro.pas.rochester.edu}}
\altaffiltext{2}{Department of Physics and Astronomy, 
                 University of Missouri, Columbia, MO 65211}
\altaffiltext{3}{Department of Earth and Environmental Sciences, University of 
                 Rochester, Rochester, NY 14627}
\altaffiltext{4}{NASA Astrobiology Institute, and Department of 
                 Physics and Astronomy, UCLA, 430 Portola Plaza, 
                 Los Angeles, CA 90095.  Current address: JPL, 
                 Caltech, Mail Stop 264-767, 4800 Oak Grove Drive, Pasadena, CA 91109}
\altaffiltext{5}{Center for Radiophysics and Space Research, 
                 Cornell University, 
                 Ithaca, NY 14853}

\begin{abstract}

Mid-infrared spectra of a few 
T Tauri stars (TTS) 
taken with the Infrared Spectrograph (IRS) 
on board the Spitzer Space Telescope 
show prominent narrow emission features 
indicating silica (crystalline silicon dioxide). 
Silica is not a major constituent of the interstellar medium; 
therefore, any silica present in the circumstellar protoplanetary disks of TTS must
be largely the result of processing of 
primitive dust material in the disks 
surrouding these stars. 
We model the silica emission features in our spectra 
using the opacities of various polymorphs of silica 
and their amorphous versions 
computed from earth-based laboratory measurements. 
This modeling indicates that the two polymorphs of silica, 
tridymite and cristobalite, which form at successively higher temperatures and low
pressures, are the dominant forms 
of silica in the TTS of our sample.  
These high temperature, low pressure polymorphs of silica 
present in protoplanetary disks are consistent with 
a grain composed mostly of tridymite named Ada 
found in the cometary dust samples 
collected from the STARDUST mission to Comet 81P/Wild 2.  The silica in these
protoplanetary disks 
may arise from incongruent melting of enstatite or from incongruent melting of
amorphous pyroxene, the latter being analogous to the former.
The high temperatures of $\sim$\,1200K-1300\,K and rapid cooling required 
to crystallize tridymite or cristobalite set constraints 
on the mechanisms that could have formed the silica in these protoplanetary 
disks, suggestive of processing of these grains during 
the transient heating events hypothesized to create chondrules.

\end{abstract}


\section{Introduction}

Silica (chemical name for the quartz group of minerals and a synonym
for silicon dioxide) is a major ingredient of the Earth's crust.  The basic
structural unit of crystalline silica is the  
tetrahedron, with the oxygen atoms at the corners 
and the silicon atom in the center.  
While in crystalline silica, structurally known as a tectosilicate, all oxygen atoms
of a given tetrahedron 
are shared with the adjacent tetrahedra forming a three dimensional network
structure, this does not hold
for most other silicates.  In silicates, SiO$_{4}^{4-}$ tetrahedra are coordinated 
by metal ions such as Mg$^{2+}$ and Fe$^{2+}$; examples of silicates are 
pyroxenes ([Mg,Fe]SiO$_{3}$) and olivines ([Mg,Fe]$_{2}$SiO$_{4}$).  
The spectra of both silica and silicate dust have vibrational
spectral features near 10$\mum$ (due to the Si--O stretching mode)
and $\simali$20$\mum$ (due to the O--Si--O bending mode).  
For silica, the main features are 
at $\simali$9, $\simali$12.6, and $\simali$20$\mum$, 
with a minor feature at $\simali$16 $\mum$ for one of 
the polymorphs of silica, cristobalite, although \citet{swdo93} note that the 16
$\mum$ feature of $\beta$-cristobalite is present at room temperature, but grows
weaker with increasing temperature until it vanishes at $\simali$ 525K.  These are
the features that we use to identify the presence of silica grains in our spectra of
protoplanetary disks, which are disks of dust and gas thought to be raw material
from which larger bodies, including planets, form.

Silica has been used to determine the temperature and pressure 
conditions at the time of formation in terrestrial rocks \citep{tb58} and also in
enstatite chondrite meteorites \citep{binns67,dodd81}.  These conditions dictate how
silica crystallizes into its various polymorphs \citep{hean94,hemley94}.  Generally, 
for pressures of 1 atm or lower, silica crystallizes as $\alpha$-quartz at 
temperatures below 846K; as $\beta$-quartz for temperatures between 846K and 1140K; 
as tridymite between 1140K and 1743K; as cristobalite between 1743K and 2000K; and 
as a liquid of silica composition above 2000K.  For a phase diagram illustrating 
these ranges of stability of silica polymorphs, see Figure 1 in Chapter 1 of 
\citet{hean94}.
At higher pressures and a wide range of temperatures, 
coesite, and then stishovite forms.  All of the crystalline forms of silica are
similar in that their bulk composition is SiO$_2$ and each of the oxygen atoms of
every SiO$_{4}^{4-}$ tetrahedron is shared with the adjacent tetrahedron; however,
the polymorphs differ in their arrangement of the SiO$_{4}^{4-}$ tetrahedra in
crystalline networks.  If silica is not allowed sufficient time to crystallize into
one of its polymorphs, it will form as amorphous SiO$_2$.
In principle, one can estimate the formation conditions 
(temperature and pressure) of the silica dust in 
astronomical environments based on the polymorph(s) 
of silica which can be inferred from
their infrared (IR) spectra. 

Silica cannot be an abundant interstellar dust species 
since the expected $\simali$9 and $\simali$12.6$\mum$ bands 
are not seen in the ISM \citep[see][]{ld02}. 
The presence of silica dust grains in the protoplanetary disk around the TTS Hen
3-600 A was indicated in its 8-13 $\mum$ spectrum \citep{honda03}.

In this work we report the detection of silica dust grains in
the disks around five T Tauri stars from the prominent emission features
characteristic of submicron silica grains.  These 
emission features are generated in the optically thin uppermost 
layers of protoplanetary disks. 
We model the spectra obtained with the Infrared 
Spectrograph \citep[IRS;][]{hou04} 
on board the {\it Spitzer} Space Telescope \citep{wer04}
with silica as a key dust component, 
with the aim to constrain the polymorph(s) 
of the silica dust in these disks and therefore the formation 
conditions (temperature and pressure history) of this dust.

\section{Data Reduction}

\subsection{Observations}

The five T Tauri stars are described in Table 1.  All were observed using both
orders of the two low spectral resolution modules of the IRS, Short-Low (SL) and
Long-Low (LL), using Staring Mode.  For more information on the spectral modules and
on Staring Mode, we refer the reader to \citet{sarg06}.  These stars were chosen for
our study of silica features because all five stars - 1RXS J161410.6-230542, IRS 49,
ROXs 42C, T51, and ZZ Tau - are SED Class II Young Stellar Objects; have prominent
features at 9, 12.6, and 20 $\mu$m wavelengths attributable to silica and often 16
$\mu$m features partly attributable to silica; and have high quality spectra with
high signal-to-noise ratios.

\subsection{Extraction and Calibration of Spectra}

We obtained basic calibrated data (BCD; flat-fielded, stray-light-corrected,
dark-current-subtracted) products from the {\it Spitzer} Science Center (SSC) for
each of the targets in our sample from the S15.3.0 IRS data calibration pipeline for
1RXS J161410.6-230542 and from the S14.0.0 IRS data calibration pipeline for IRS 49,
ROXs 42C, T51, ZZ Tau.  First, we identified bad pixels as described by
\citet{wat07} in their explanatory supplement.  In addition, we also identified as
bad pixels a string of pixels with large values of e-/sec in the BCD data for the
first nod of the first order SL observation of ZZ Tau from Campaign 4.  Some of the
affected pixels were flagged as ``radhit'' detections (most likely a cosmic ray hit)
in the associated bmask; we marked as bad pixels these and additional neighboring
pixels that also appeared to be affected by the cosmic ray.  As 12.5 $\mu$m is about
the peak wavelength of a key diagnostic silica feature, fixing these pixels was
critical for our purposes of analyzing dust emission features of silica.  We correct 
for bad 
pixels by interpolation of neighboring pixels in the spectral direction, as
described by \citet{wat07}.  First, we take the average of the pixel one pixel
above and one pixel below.  This is applied to all bad pixels in the bcd data.  If
a given pixel is not fixed, we then take another iteration; this time, we take the
average of the pixel one pixel above and two pixels below the bad pixel.  This
process is repeated, with subsequent iterations going two pixels above and one
pixel below, then two pixels above and two below, then finally one more iteration
going one pixel above and one below.

The spectra presented in this paper were extracted from the BCD data using the
Spectral Modeling, Analysis, and Reduction Tool \citep[SMART;][]{hig04} via methods
described by \citet{sarg06} and \citet{fur06}.  For all targets in our sample, we
subtracted sky for SL and LL by subtracting the pixel-fixed BCD data for the same
order but other nod position.  We extracted a spectrum each DCE separately, then
used the extracted spectrum from each DCE to compute, first, the average spectrum
and, second, its uncertainties from the standard deviation of the mean.  Wavelength
calibration of our data, tapered column extraction for each order of SL and LL, and
flux density calibration of spectra using Relative Spectral Response Functions
(RSRFs) was achieved as described by \citet{sarg06}.  The same high spectral
resolution template of $\alpha$ Lac (A1\,V; M. Cohen 2004, private communication)
used by \citet{sarg06} was used to generate RSRFs for SL and LL flux calibration for
IRS 49, 
ROXs 42C, T51, and ZZ Tau, and for SL flux calibration for 1RXS J161410.6-230542. 
For the LL flux calibration of 1RXS J161410.6-230542, we applied LL second order, LL
bonus order, and LL first order (from its short-wavelength end to 36 $\mu$m
wavelength) RSRFs generated from data of $\xi$ Dra and the template for $\xi$ Dra
from \citet{coh03}; past 36 $\mu$m, we applied a LL first order RSRF generated from
data of Markarian 231 and the template for Markarian 231 \citep[J. Marshall,
private communication;][]{marsh07,armus07}.

We note the Humphreys-$\alpha$ and [Ne II] lines at 12.37 and 12.81 $\mu$m,
respectively, are located near the central wavelengths of the two narrow features
consituting a double peak in the opacity profile of $\alpha$-quartz (see discussion
later in Section 3.1).  \citet{sarg06} concluded $\alpha$-quartz was present in the
IRS spectrum of TW Hya based upon two narrow emission features located at the
wavelengths of the two peaks of the $\alpha$-quartz opacity profile used in that
study.  Upon detailed inspection of the Short-High spectrum of TW Hya presented by
\citet{uch04}, we find unresolved emission lines from Humphreys-$\alpha$ and [Ne II]
and no underlying emission from the $\alpha$-quartz double feature.  We have looked
at the Short-High spectra from Staring Mode observations of 1RXS J161410.6-230542
(AOR ID\# 5453824) and ROXs 42C (AOR ID\# 6369792) for these lines.  The AORs had
two DCEs per nod position and two nod positions; in these Short-High data, we found
no emission from Humphreys-$\alpha$ or [Ne II].

\subsection{Extinction Correction of spectra}

For all objects except the highly extinguished ($A_{V}$=10.7) IRS 49, we corrected
our RSRF-calibrated spectra in the same manner as described by \citet{sarg06},
having adopted values of extinction at V-band, $A_{V}$, as listed in Table 1.  For
IRS 49, extinction correction was accomplished via the same method, and also using a
lower $\tau_{9.7}$ assuming $A_{V}/\tau_{9.7}$=25, a value suggested for extinction
in dense star forming regions like Ophiuchus (in which IRS 49 resides) by
\citet{chiar07}.

\subsection{Mispointing}

We account for mispointing along the slits (the cross-dispersion direction) by
adjusting the positions of our 3-5 pixel wide extraction boxes.  The degree of
mispointing is typically much less than 1 pixel (1.8 $\arcsec$ in SL), and we
adequately account for this by using the BCD flatfielding.  Mispointing in the
dispersion direction is harder to detect and causes a loss of signal at every
wavelength.  We assume that the flux density from one nod being systematically lower
than that from the other nod was due to loss of signal from dispersion-direction
mispointing.  To correct for this signal loss, we multiplied the nod lower in flux
density by a scalar to minimize the squared differences in flux at each wavelength
in the order.  For more about correction for mispointing, see the Appendix.

\subsection{Uncertainties}

Uncertainty in flux density was determined by the standard deviation from the mean
for the scalar-multiplied spectra from all DCEs.  For IRS 49, ROXs 42C, and T51,
there were only 2 DCEs (the 2 nod positions), so for these the standard deviation
from the mean equals half the difference of the spectra at the 2 nod positions. 
Also, for the reasons listed by \citet{sarg06}, we set relative uncertainties lower
than 1\% to 1\%.

\subsection{Photosphere subtraction}

Three of the five objects whose spectra are analyzed in this study, 1RXS 
J161410.6-230542, ROXs 42C, and ZZ Tau, have Spectral Energy Distributions (SEDs)
indicative of very little mass in small dust grains in the innermost disk regions. 
The SEDs of these three objects \citep{sil06,mcc08,fur06} have optical and JHK
photometry that is consistent with stellar photospheric emission, with ZZ Tau
photospheric even to L band ($\sim$3.5 $\mu$m).  Also in each of these three cases,
the excess above stellar photosphere as seen from the IRS spectrum at 5 $\mum$
wavelength is very small.  Therefore, photosphere subtraction is important for these
spectra in order to isolate the emission from circumstellar dust
\citep[see][]{sarg06} at the short wavelength end of the IRS spectra.  The low
extinction to these three objects (see Table 1) gives us confidence that we can
reasonably account for stellar photosphere emission in the IRS spectra.  Therefore,
for each we subtract a Planck function 
at the stellar effective temperature scaled to equal the flux at the dereddened J
band flux of each object.  At {\it Spitzer} IRS wavelengths ($>$ 5 $\mum$), we are
effectively subtracting the Rayleigh-Jeans tail of these Planck functions.  The
solid angles and temperatures used for the Planck functions representing stellar
photosphere emission are listed in Table 1.  For the stars T51 and IRS 49, there is
significant excess at all IRS wavelengths, so no stellar continuum was subtracted.

\section{Analysis}

\subsection{Silica Opacities}

\Citet{sk61} and \citet{wc96} 
measured the complex dielectric function 
$\varepsilon(\lambda) = \varepsilon_1(\lambda) + i \varepsilon_2(\lambda)$
for $\alpha$-quartz at ``room temperature'' ($\simali$300\,K); 
\citet{gp75} obtained $\varepsilon(\lambda)$ for $\alpha$-quartz 
at 295\,K, 505\,K, 620\,K, 785\,K, and 825\,K 
as well as for $\beta$-quartz at 860\,K and 975\,K. 

To compare these dielectric functions,  
we calculate, for the sake of illustration, 
the opacities for $\alpha$-quartz
with the same grain shape distributions of 
either CDE \citep{bh83} or CDE2 \citep{fab01},
using the dielectric functions from 
\citet{sk61} for quartz at $\simali$300\,K, 
\citet{wc96} for quartz at $\simali$300\,K, 
and \citet{gp75} for quartz at 295\,K
and taking the ``$\frac{2}{3}-\frac{1}{3}$-approximation'' 
for the ordinary and extraordinary rays 
\citep[see \S\,A in ][]{sarg06}.  
We find that the opacities computed from the three sets of dielectric functions
are nearly identical.  
This implies that we can have confidence
not only in the optical properties of $\alpha$-quartz
at room temperature, but also in the dielectric functions
of $\alpha$-quartz and $\beta$-quartz 
reported by \citet{gp75} at other temperatures.

We show opacity curves for CDE and CDE2 shape distributions 
using the dielectric function from \citet{gp75} for 
$\alpha$-quartz at 295\,K in Figure 1.  We take the mass density of $\alpha$-quartz 
to be 2.65$\g\cm^{-3}$ \citep{hean94}.  In the literature, the CDE and CDE2 grain
shape distributions 
have been widely considered to explore the effects 
of dust shapes on the opacity profiles.  Both CDE and CDE2 assume that the size of
the grains under consideration are in the Rayleigh limit; i.e., 2$\pi$a/$\lambda$
$<<$ 1.  Thus, the grain sizes for CDE and CDE2 shape distributions used in our
models of mid-infrared emission are submicron.  The CDE shape distribution
\citep{bh83} assumes a distribution of ellipsoidal shapes in which each shape,
defined by the axial ratios of the ellipsoids, is statistically weighted equally. 
The CDE2 shape distribution \citep{fab01} assumes a distribution of ellipsoidal
shapes in which ellipsoids that are almost spherical are weighted more heavily than
more extreme (highly elongated along one or two axes or highly flattened along the
other axes or both) shapes; the most extreme shapes in CDE2 are given zero weight. 
CDE is therefore more heavily weighted toward very nonspherical ellipsoids than
CDE2.  Our results confirm what others have shown in the literature \citep[e.g. see
the opacities of spheres, CDE, and CDE2 for 
forsterite by ][]{fab01} -- shape distributions of 
ellipsoids that more heavily weight toward very nonspherical 
ellipsoids with respect to spheres result in opacity profiles 
with features centered at increasingly longer wavelengths.  

To approximate the effects of quartz of 
a range of temperatures, 
in Figure 2 we show the $\beta$-quartz opacity profile 
obtained by averaging the opacity curves of $\beta$-quartz 
of 975\,K and 860\,K.  We take the mass density of $\beta$-quartz to be
2.53$\g\cm^{-3}$ \citep{hean94}.  Similarly, we obtain the high-temperature
$\alpha$-quartz profile
from averaging the opacity of $\alpha$-quartz at 825\,K and 785\,K, and the
lower-temperature $\alpha$-quartz profile from averaging the opacity of
$\alpha$-quartz of 620\,K and 505\,K, all with a CDE shape distribution.
Note the progression of the 12.6$\mum$ feature complex 
from double-peaked to single-peaked 
as the quartz temperature increases (see Fig.\,2).  We assume that a given polymorph
of silica, once formed, 
can exist stably at temperatures and pressures within its stability field on the
phase diagram or metastably outside the stability field at lower temperatures if the
physical conditions change relatively rapidly.  However, we assume a polymorph of
silica becomes 
another polymorph if subjected to temperatures 
higher than allowed by its stability
field in the silica phase diagram.

Infrared transmission spectroscopy by \citet{ply70} 
and absorption measurements by \citet{wi93} for coesite 
and stishovite rule out these higher 
pressure polymorphs of silica in our sample of TTS disks.  For the intermediate
pressure polymorph, coesite, 
both \citet{ply70} and \citet{wi93} show four features 
with widths similar to that of the 12.6$\mum$ feature 
between 14 and 23$\mum$, but all with strengths greater 
than that of the 12.6$\mum$ feature -- one feature near 14$\mum$, 
a pair of slightly stronger features at $\simali$16--17$\mum$, 
and the strongest of the four features at $\simali$20--23$\mum$.  The spectra with
prominent silica features 
do often have a feature at $\simali$16--17$\mum$, 
but it is single, and, in addition, such spectra 
do not have a narrow feature at 14$\mum$ of greater prominence 
than the 12.6$\mum$ feature.  This rules out coesite.  Stishovite can be more easily
ruled out, as both \citet{ply70} and \citet{wi93} show stishovite to lack a 9$\mum$
feature, while the spectra of the five silica exemplars analyzed in this study have
prominent 9 $\mum$ features.

To represent the two higher temperature, low pressure 
polymorphs of silica, tridymite and cristobalite, 
we use the opacity obtained from the transmission 
measurements of submicron annealed silica grains 
embedded in Potassium Bromide, KBr, by \citet{fab00}.  
This silica was formed by heating amorphous SiO$_2$ grains 
at 1220\,K for 5 hours, and was found to be mostly 
cristobalite but also partially tridymite.  
This is supported by the fact that the cristobalite 
presented by \citet{ply70} and \citet{rok98} have 
a 16$\mum$ feature of prominence similar to but slightly 
less than that of the 12.6$\mum$ feature, 
whereas for tridymite the 16$\mum$ feature is 
almost nonexistent (see the tridymite absorbance 
shown by \citet{hof92}).  
The annealed silica presented by \citet{fab00} 
has a 16$\mum$ feature of prominence closer to 
that of ``low cristobalite'' as measured by \citet{rok98}.  
We also note that the opacity profile of the silica annealed 
by \citet{fab00} at 1220\,K for 5 hours bears a close resemblance 
to the IR spectrum of the products of annealing for 120 hours 
at 1300\,K of silica formed from iron-bearing silicates by \citet{hnd98}.  
We present in Figure 3 the opacity of the annealed silica 
used in our models.

For amorphous SiO$_2$, the dielectric functions 
at mid-IR wavelengths were measured by \citet{hm97}, 
\citet{steyer74}, and \citet{koike89}.  
In our dust modeling we adopt the dielectric functions 
of amorphous SiO$_2$ at 300\,K of \citet{hm97}.  
In addition, we consider obsidian, as measured by \citet{koike89}. 
To compute the opacities of obsidian and amorphous SiO$_2$, 
we take their mass densities to be 
2.384$\g\cm^{-3}$ \citep{koike87} 
and 2.21$\g\cm^{-3}$ \citep{fab00}, respectively.  
We plot in Figure 3 also 
the opacities of obsidian and amorphous SiO$_2$ in the CDE 
shape distribution.  
The shape distribution affects 
the opacities of obsidian and amorphous SiO$_2$ less than 
it does for $\alpha$- and $\beta$-quartz; 
for this reason, we only use the opacities of 
obsidian and amorphous SiO$_2$ with a CDE shape distribution, 
and we do not plot the opacities of these grains in the CDE2 
shape distributions.

\subsection{The KBr Effect}

Embedding small grains in KBr is commonly 
used to measure the grain opacity in the IR.  
This technique has limitations -- optical constants 
are not directly measured, so it is not possible to 
explore grain shapes, porosity, or size.  
The embedded grains may clump together, 
reducing the contrast in sharp features.  
Moreover, the index of refraction of KBr ($\simali$1.5) 
can distort strong resonance 
features in the spectrum from how they would appear in a spectrum of the same grains
in a vacuum (i.e., in a protoplanetary disk).

Annealed silica and crystalline pyroxene are 
the only grain components in our models which 
depend on a KBr-embedded measurement.  
We model the KBr effect on $\beta$-quartz (see Figure 4) 
to illustrate the probable effects of KBr 
on our annealed silica opacities.  
We compute the opacities of solid, small $\beta$-quartz grains 
in a KBr medium and in vacuum 
for both CDE and CDE2 shape distributions.  
We use the optical constants of $\beta$-quartz at 975K computed by \citet{gp75} as
the optical properties of silica grains that are embedded in a KBr medium whose
6--28$\mum$ optical constants we take from \citet{hea71}.
As shown in Figure 4, for the CDE shape distribution, 
the peak wavelengths of the 9, 12.6, and 20$\mum$ features 
are little altered by embedding in KBr. 
The peak wavelength of the 9$\mum$ feature for $\beta$-quartz 
in the CDE2 shape distribution embedded in KBr is at a slightly longer wavelength
than that for $\beta$-quartz grains in the CDE2 distribution in vacuum.
This also holds for the 20$\mum$ feature.  
The peak wavelength of the 12.6$\mum$ and 16 $\mum$ features 
are much less affected by either being in KBr 
or with a different shape distribution.  

Note in Figure 4 that if the shape distribution 
for the dust in KBr is already at an extreme (e.g. CDE), 
models with the same shape distribution but in vacuum 
will have spectral features not much different
 (in width and central wavelength) from those in KBr.  
We note that all opacities of $\beta$-quartz are somewhat 
higher for dust in a KBr medium, also found by \citet{jag98}.

\subsection{Other Dust Opacities}

For submicron amorphous silicates, we compute 
opacities for the CDE2 shape distribution \citep{fab01}.  We take the dielectric
functions of MgFeSiO$_{4}$ 
of \citet{dor95} for amorphous olivine.
For amorphous pyroxene we use the dielectric functions 
of Mg$_{0.7}$Fe$_{0.3}$SiO$_{3}$ from \citet{dor95}.  We take the mass densities of
both olivine and pyroxene
to be 3.3$\g\cm^{-3}$.

To account for the effects of grain growth, 
we consider porous dust.
Porous material is simulated by applying 
the Bruggeman effective medium theory \citep{bh83}, 
assuming the porous grains are 60\% vacuum by volume.
The effective dielectric functions 
for porous amorphous pyroxene and olivine are then computed.
Mie theory \citep{bh83} is then used to compute 
the opacities of spherical porous grains of 
radii of 5$\mum$.
As noted by \citet{sarg06}, 
the opacity of 5$\mum$-radius, 60\%-porous 
amorphous silicate grains is similar to that 
for 2$\mum$-radius solid dust, 
the size of dust used by \citet{bouw01} to investigate 
grain growth in Herbig Ae/Be disks.

We also include pyroxene (crystalline) and forsterite 
(of which the density is also 3.3$\g\cm^{-3}$) in our analysis.  
For pyroxene, we use the opacity of ``En90'' \citep{chi02}. 
With a stoichiometric composition of Mg$_{0.9}$Fe$_{0.1}$SiO$_{3}$, 
``En90'' is quite effective in fitting the pyroxene features 
at 9.3, 10.5, 11.2, and 11.6$\mum$ within the 10$\mum$ complex 
of FN Tau \citep[see][]{sarg06}.  
For forsterite, we use the optical constants of \citet{sog06}
which are similar to that of \citet{sp73}, 
to compute the opacity for a distribution of ellipsoidal shapes.

Using the formalism of \citet{bh83}, we have constructed 
an opacity profile for a distribution of forsterite grains 
in which all ellipsoidal axial ratios are given an equal weight, 
as in CDE.  However, we exclude the most extreme ellipsoidal axial 
ratios in the following manner.  

Our shape distribution, which we call ``tCDE'' 
(truncated Continuous Distribution of Ellipsoids), 
is that we confine $L_1$ and $L_2$ (which are the $L_j$ parameters described by
\citet{bh83}) within the triangle 
in the $L_1$-$L_2$-space specified by vertices 
($L_1=0.1$, $L_2=0.05$), (0.1, 0.895), and (0.99, 0.005).  
For CDE, all shapes are allowed: the triangle specified 
by the verticies (0,0), (0,1), and (1,0).  
The opacity profile of forsterite for CDE has a feature 
peaking at 11.3$\mum$ which is close to the feature seen 
in astronomical data (but often peaking at a slightly 
longer wavelength). The opacity profile for tCDE ``rounds'' 
this peak somewhat, pushing the peak toward slightly shorter 
wavelengths, and this results in marginally better fits to 
astronomical spectra.

We will discuss ``tCDE'' at a greater length in \citet{sarg08}, 
but here we justify this choice by showing in Figure 5 
our model fits to ROXs 42C with forsterite grains having
either a CDE shape distribution (``CDE'' model)
or a tCDE distribution (``tCDE'' model). 
The former (CDE) has $\chi^{2}$ per degree of freedom 
(d.o.f.) of $\simali$5.1, while for the latter (tCDE) 
$\chi^{2}/{\rm d.o.f.} \approx 4.3$ which is somewhat better; 
the tCDE model does a slightly better job in fitting 
the 11 and 23$\mum$ complexes of the spectrum of ROXs 42C 
but is overall very similar to CDE.  
\citet{sarg06} found great 
similarity between the opacity of forsterite and silica grains 
with a CDE shape distribution and that of porous dust
with a porosity of 60\%.  Thus the effects of shape distribution and porosity cannot
be clearly distinguished in the spectra.

No hint of Polycyclic Aromatic Hydrocarbon (PAH) emission is seen 
in our spectra, except possibly ZZ Tau at 6.2 $\mum$.  Especially 
telling is the lack of the ubiquitous 6.2, 7.7$\mum$ PAH features, 
which are clearly distinguishable from silicate or silica emission; 
in ZZ Tau, no 11.3 $\mum$ PAH emission is evident.

We note that the opacity profiles of some solids 
are similar to that of silica, but nevertheless 
can be ruled out.  
\citet{kn07} measured the transmittance 
of H$_{2}$Si$_{2}$O$_{4}$ and Si$_{2}$O$_{3}$. 
They found in some of their spectra the features 
peak at wavelengths very close to those of silica 
($\simali$9, 12.6, and 21.5$\mum$); 
however, many of their spectra also show a narrow 
feature at 11.36$\mum$ with a width similar to 
their 12.6$\mum$ feature but of greater prominence.  
This feature is not seen in the spectrum of ZZ Tau.  
In the spectra of 1RXS J161410.6-230542 and ROXs 42C 
there is a feature near 11.1$\mum$, 
but as will be shown in our model fits to their spectra, 
this peak can be attributed to forsterite.  
Also, the opacity curves computed using the optical constants of
Na-bearing amorphous aluminosilicates determined by \citet{mut98} 
exhibit features near 9$\mum$ and 21$\mum$ 
similar to those of silica of various versions; 
however, the Na-bearing aluminosilicates 
lack the distinctive 12.6$\mum$ feature which 
is seen in the emission spectra of the 5 T Tauri stars 
and in the opacity spectra of all versions of silica.

\subsection{Models}

We model the spectra of each of five T Tauri stars as a sum of featureless continuum
emission 
(from optically thick disk midplane, iron or amorphous carbon grains, 
and very large $>$10$\mum$ silicate grains) and optically 
thin emission from dust grains with strong infrared resonances in the disk 
atmosphere \citep[see][]{sarg06}.  It is not our goal to model self-consistently 
the entire spectral energy distribution from the near-IR to submm; 
such would require radiative transfer models 
such as those of \citet{cal92} and \citet{daless01}.

We reject silicate self-absorption, a mix of emission and absorption by similar 
kinds of dust but with cooler dust in front of warmer dust, 
as an explanation of the 9 and 12.6 $\mum$ features we see 
in our spectra.  Self-absorption in the 10 $\mum$ silicate complex would 
give apparent emission features in the 8 and 12 $\mum$ wings of the 10 
$\mum$ complex; however, such would require the optical depth of the cool 
amorphous silicate dust to exceed 1.  Since amorphous silicate 
dust in the ISM gives $A_{V}/\tau_{9.7}$ of 18 or 25 (see \S2.3), this would 
mean the intervening dust has $A_{V}$ of 18 to 25.  Such dust would not be 
at the disk surface; otherwise it would heat up to the same temperature as 
surrounding dust and not give rise to an absorption feature.  Such dust would 
have to be further away from the disk and star, and therefore would also be 
in front of the star in the line-of-sight to the star, and we would therefore 
find $A_{V}$ of 18 to 25 to all stars with what appear to be prominent 9 and 12.6 
$\mum$ silica features.  All of our objects have $A_{V}$ less than 18, the minimum 
$A_{V}$ required to result in self-absorption.

We approximate the IR emission from dust 
with a large range of temperatures by two temperatures.  
A similar approach was employed by \citet{kast06} 
to model the IRS emission spectra of the dust around 
the hypergiant R66 in the Large Magellanic Cloud 
and by \citet{chen06} to model that of five debris disks 
(see especially the two dust temperatures used 
in the model for $\eta$ Tel).  We assume that, for a given protoplanetary disk, 
all dust grains, regardless of composition, within the same arbitrarily small volume 
of the part of the disk atmosphere giving rise to optically thin silicate and silica 
emission are at the same temperature.  \citet{kadull04} find in their disk models 
that, in a layer above the layer giving rise to dust emission features, the gas and 
dust temperatures are within 10\% of each other.  As gas density increases closer 
towards the disk midplane, the gas and dust temperatures in the layer giving rise to 
dust emission features should be even closer to each other.  The dense gas 
effectively should equalize the temperatures of dust grains of different optical 
properties (resulting from, e.g., different compositions, different sizes).  Our
two-temperature 
models therefore model dust composition for two disk regions: the warmer inner disk 
regions and the cooler outer disk regions.

Our model is a sum of blackbodies 
at two temperatures plus optically thin emission 
from dust at those temperatures.  
Mathematically, these models are given by
\begin{eqnarray}
F_{\nu}(\lambda)^{\rm mod} & = & 
       B_{\nu}(\lambda,T_{c}) \left[{\Omega}_{c} + 
       \sum_i a_{c,i}\kappa_{i}({\lambda})\right] + \nonumber \\
& & B_{\nu}(\lambda,T_{w})
\left[{\Omega}_{w}+\sum_j a_{w,j}\kappa_{j}({\lambda})\right] ~~,
\end{eqnarray}
where $F_{\nu}(\lambda)^{\rm mod}$ is the model flux density; 
$B_{\nu}(\lambda,T)$ is the Planck function at temperature $T$ 
and wavelength $\lambda$; $T_{c}$ is the temperature of 
``cool'' dust and $T_{w}$ is the temperature of ``warm'' dust; 
$\Omega_{c}$ ($\Omega_{w}$) is the solid angle of 
a blackbody representing the continuum emission from 
cool (warm) dust; 
$a_{c,i}$ ($a_{w,i}$), the mass weight for cool (warm) dust, 
equals $m_{c,i}/d^2$ ($m_{w,i}/d^2$),
where $m_{c,i}$ ($m_{w,j}$)
is the mass of dust species $i$ 
at $T_c$ ($T_w$) and $d$ 
is the distance to the T Tauri star 
(see Table 1 for the stellar distances).  We specify a grid of temperature pairs,
$T_{c}$ and $T_{w}$, 
not allowing either of the temperatures to exceed 1401\,K, 
as this temperature is intermediate between the stability 
limits of forsterite and enstatite of 1430\,K and 1370\,K, 
respectively \citep{posch07}.  For a given temperature pair, 
our model equation is linear in solid angles and optically thin dust mass weights. 
We minimize 

\begin{eqnarray}
\chi^{2} & = & \sum_k 
\left[\frac{F_{\nu}(\lambda_k)^{\rm irs}
 - F_{\nu}(\lambda_k)^{\rm mod}}{\Delta F_{\nu}(\lambda_k)^{\rm irs}}\right]^{2} ~~,
\end{eqnarray}
with respect to each of the solid angles 
and mass weights, where $F_{\nu}(\lambda_k)^{\rm irs}$ is the observed flux density 
at the $k$-th waveband and $\Delta F_{\nu}(\lambda_k)^{\rm irs}$ is the uncertainty
in the observed flux density at the $k$-th waveband.
If any of the mass weights computed from an iteration of $\chi^{2}$ minimization are
negative, 
the component with the most negative integrated flux from 7.7-23 $\mum$ is set to
zero, 
and we then minimize $\chi^{2}$ again.  
This process is iterated until there are no components 
with negative mass weights.  
The solution found in the iteration for which no components 
are negative is the best fit for a given temperature pair. 
The process is carried out for another temperature pair.  
The global best fit for a spectrum is 
at the temperature pair that minimizes $\chi^{2}$ 
per degree of freedom.  In all of our best fits, the silica component at both low
and high 
temperatures was positive.  The number of degrees of freedom 
is equal to the number of data points minus four (since we consider two dust
temperatures and two solid angles) minus the number of mass weights at both
temperatures altogether.

Figure 5 is a sample fit of a model with
annealed silica to the 7.7--37$\mum$ IRS spectrum of ROXs 42C (all other models,
including the other models of ROXs 42C, discussed or shown in this paper are fit
over 7.7-23 $\mum$).  It is seen that our two-temperature model fits the spectrum of
ROXs 42C over almost the entire wavelength span of the IRS.  What interests us most
in this study are the spectral features diagnostic of silica, 
the 9, 12.6, 16, and 20$\mum$ features.  
For this reason, we minimize $\chi^{2}$ 
over the smallest range of wavelengths sufficient 
to adequately cover these features, 7.7 to 23$\mum$.  
We determine a very conservative estimate of the uncertainty for each nonzero 
dust component and solid angle in the model 
as the amount that resulted from increasing $\chi^{2}$ per degree of freedom by
$\simali$ 1 from the best-fit value.  
The models shown in all figures after Figure 5 and whose parameters 
are summarized in Table 2 are constructed by minimizing $\chi^{2}$ over 7.7--23$\mum$.

We obtained models of quartz in each of the CDE and CDE2 
shape distributions (where the same shape distribution is used for quartz at the low
temperature, $T_{c}$, as for quartz at the high temperature, $T_{w}$).  We used the
opacity of the 295\,K $\alpha$-quartz 
for the ``cool'' silica component.  For the opacity of the ``warm'' silica component, 
we use the $\beta$-quartz, high-temperature $\alpha$-quartz, and lower-temperature
$\alpha$-quartz opacity curves described in \S3.1 in reference to Figure 2.
The opacities computed from the dielectric constants of $\alpha$-quartz at 505\,K
and 620\,K were averaged to give an ``average 562.5\,K $\alpha$-quartz opacity'';
785\,K and 825\,K $\alpha$-quartz opacities were averaged to give an ``average
805\,K $\alpha$-quartz opacity''; and 860\,K and 975\,K $\beta$-quartz opacities
were averaged to give an ``average 917.5\,K $\beta$-quartz opacity''.
The model with the lowest difference between 
$T_{w}$ and ``average $\alpha$- or $\beta$-quartz opacity temperature'' (562.5\,K,
805\,K, or 917.5\,K) is the best fit model for quartz grains in the shape
distribution under consideration (only CDE and CDE2 were considered).  A model is
rejected if $T_{w}$ is greater than the highest temperature allowed according to the
stability field 
for a given silica polymorph (for silica polymorph stability fields, see Section 1).

\section{Results}

\subsection{ROXs 42C}

ROXs 42C was identified by \citet{ba92} as the easternmost stellar source within the
positional error of the X-ray source ROX 42 found by the Einstein satellite
\citep{mont83}.  \citet{math89} found it to have a double-lined spectroscopic binary
with a 36 day period, while \citet{ghez93} found it also to be a visual binary
separated by $\simali$ 0.16\arcsec; \citet{jm97} conclude that ROXs 42C is a
hierarchical triple - a spectroscopic binary orbited by single star further out. 
\citet{ratz05} agree with the assessment that ROXs 42C is a hierarchical triple, but
report that the visual binary is now separated by 0.277\arcsec.  Assuming a distance
to Ophiuchus of 140pc (see Table 1), this implies a separation of $\simali$ 39 AU
between the double-lined spectroscopic binary and the single star.  \citet{jensen96}
calculate the separation of the double-lined spectroscopic binary to be 0.27 AU.  We
note the similarity between ROXs 42C and Hen 3-600, itself a spectroscopic binary wi
 th another star at 70 AU orbiting the binary.  We also note that the disk in the
Hen 3-600 system also has abundant silica as indicated by its IR spectrum
\citep{honda03,uch04}.  The state of accretion in the disk(s) in this system is
unknown, as \citet{ba92} report the Equivalent Width of H-$\alpha$ to be neither in
absorption or emission, but such that the photospheric line was filled in by
1.6\Angstrom\, of emission.  However, because of the multiple stars in this system,
and because it is not known where the disk(s) is (are) located, we do not know
whether active accretion could be occurring currently onto one star and its
spectral signature is obscured, or whether there is no active accretion in this
system.

The IRS spectrum of ROXs 42C has prominent narrow features at 9 and 12.6 $\mu$m from
silica, in addition to features around 11, 23, 28, and 33.5 $\mu$m identified with
forsterite and pyroxene.  We show the model fits to ROXs 42C using each of annealed
silica, obsidian, amorphous SiO$_2$, and quartz grains in the CDE and CDE2 shape
distributions in Figure 6.

We rule out the models using the amorphous forms of silica - obsidian and amorphous
SiO$_2$ - because the 12.3-12.4 $\mu$m features of these two silicas are of much
less strength with respect to their 9 $\mu$m features than for the crystalline forms
of silica.  This translates into being able to fit the 9 and 20 $\mu$m features
somewhat reasonably but not having enough strength in the 12.3-12.4 $\mu$m feature
to fit the data.  In addition, the 12.3-12.4 $\mu$m feature from the amorphous
versions of silica is centered shortward of the feature in the data by $\sim$ 0.2
$\mu$m and is noticeably wider, with a Full Width Half Maximum (FWHM) of $\sim$ 1
$\mu$m, than that in the data (with a FWHM $\sim$ 0.5 $\mu$m).  The $\chi^{2}$ per
degree of freedom for the model using obsidian is 5.4; for the model using amorphous
SiO$_2$, it is 5.2.

Next, consider $\alpha$-quartz, $\beta$-quartz, tridymite and cristobalite
(tridymite and cristobalite being represented together by annealed silica).  For
quartz in the CDE2 shape distribution, the lowest difference between model $T_{w}$
and ``average opacity temperature'' (temperature at which quartz optical constants
were measured) happened for $\beta$-quartz.  A $\chi^{2}$ per degree of freedom of
4.5 was obtained for this model.  The fit to the 9 $\mu$m feature is good, coming
from a combination of $\beta$-quartz and pyroxene.  Also, the model is within the
error bars of a little over half the data points between 20.5 and 22 $\mu$m. 
However, the spectrum of ROXs 42C lacks the double feature at 12.6 $\mum$ in the
model, and the model does not have a strong 16 $\mu$m feature like ROXs 42C. 
Forsterite gives a 16 $\mu$m feature, but forsterite is insufficient by itself to
fit the 16 $\mu$m feature in ROXs 42C.

A $\chi^{2}$ per degree of freedom of 5.0 was obtained for the model using quartz in
the CDE shape distribution.  The fit to the 12.6 $\mu$m feature is improved due to
the feature being single-peaked in higher-temperature $\alpha$-quartz.  However,
this model provides a poor fit to 20.5 - 22.0 $\mu$m.  The model feature at 9 $\mu$m
peaks slightly longward of the feature in the data.  A shape distribution
intermediate between CDE and CDE2 could fit the 9 $\mu$m feature, but would still
provide a 20 $\mu$m feature that would overshoot the ROXs 42C spectrum between 20.5
and 22.0 $\mu$m.  Furthermore, the model does not have a 16 $\mu$m feature sharp
enough to fit the 16 $\mu$m feature in the data.  We believe this definitely rules
out $\beta$-quartz as a candidate to explain the silica features of ROXs 42C;
$\alpha$-quartz at temperatures of 505K-620K is a possibility to explain the
features, but we believe it to be of lower probability.

This leaves annealed silica.  We show the model using annealed silica at low and
high model temperatures in Figure 6.  A $\chi^{2}$ per degree of freedom of 3.2 was
obtained for this model, the lowest for any of the models of ROXs 42C.  The fit to
the 9 $\mu$m, 12.6 $\mu$m, and 16 $\mu$m features and the 20.5-22.0 $\mu$m region is
the best of all models fit to ROXs 42C, and this is reflected in the $\chi^{2}$ per
degree of freedom (Table 2 and Figure 7).  We note the opacity of annealed silica
that we use comes from KBr pellet transmission measurements, but, as discussed in
Section 3.2, we expect that the opacity curve for the annealed silica grains in KBr
would be very similar in terms of spectral feature central wavelength and width to
the opacity in vacuum.  We conclude that annealed silica best matches the spectrum
of ROXs 42C.

\subsection{ZZ Tau}

\citet{sim96} found ZZ Tau to be a binary system with a separation of stars that
increased from 0.0338\arcsec\, in 1994 to 0.0420\arcsec\, in 1996.  Subsequent
measurements of the positions of the stars in the binary let \citet{schaef06} solve
for the orbital parameters of this system.  \citet{schaef06} compute the semi-major
axis expressed as an angle of 0.061\arcsec, implying a separation of 8.5 AU. 
\citet{ps97} report EWH$\alpha$ from this system of 15 \Angstrom\, and suggest that
this system is either a hybrid between a weak-lined and a classical T Tauri Star
(the disk around one star actively accreting, the disk around the other not) or that
one of the two stars in this system has an inner gap (a ``transitional disk'').

ZZ Tau was also fit by amorphous SiO$_2$, obsidian, quartz in both CDE and CDE2
shape distributions, and annealed silica.  For this object and for the three other
objects, we only show the best fit.  As with ROXs 42C, amorphous SiO$_2$ and
obsidian gave features much too weak, too wide, and centered at the wrong wavelength
to fit the 12.6 $\mu$m feature of ZZ Tau.  The fits using quartz in both shape
distributions were rejected because the high temperature in the model was beyond the
range of stability for $\beta$-quartz.  The fit to ZZ Tau using annealed silica
(Figure 8, top) fit quite well at 9, 12.6, 16, and 20 $\mu$m, and it also proved to
be the best fit in terms of $\chi^{2}$ per degree of freedom, which amounts to 3.8
(see Table 2 and Figure 7).

\subsection{1RXS J161410.6-230542}

Of the 7 sources reported by \citet{metch06} to be located within 10\arcsec of 1RXS 
J161410.6-230542, only one was confirmed to be associated with this star.  This
companion is 0.222\arcsec\, separated from the primary and is 0.21 magnitudes
fainter at K-band \citep{metch06}.  At an assumed distance to this object of 145pc
(Table 1), this implies a separation of 32 AU between the components. 
\citet{prei98} report EWH$\alpha$ from 1RXS J161410.6-230542 of 0.96\Angstrom, which,
when combined with the K0 spectral type assigned to this system by \citet{pas07},
suggests little or no accretion is occurring from the disk to the star in the 1RXS 
J161410.6-230542 system.

Annealed silica provided the best fit to the {\it Spitzer} IRS spectrum of 1RXS 
J161410.6-230542.  As seen at the bottom of Figure 8, the fits to the 9, 12.6, 16,
and 20 $\mu$m features are all adequate with our model.  The $\chi^{2}$ per degree
of freedom is 6.6 for this model.  Of all the versions of silica discussed in this
paper, only annealed silica has a single, isolated peak at 16 $\mu$m.  The 16 $\mu$m
feature in our model is coming equally from forsterite and cold annealed silica, and
the model fits the narrow ($\sim$ 1 $\mu$m FWHM) peaks in the spectrum fairly well
at 11 and 19 $\mum$, which are features associated with forsterite, meaning the fit
to the 16 $\mu$m feature would be worse without the contribution from annealed
silica.

We note that the curious trapezoidal shape of the 12.6 $\mu$m feature does not
perfectly match the rounded shape of the 12.6 $\mu$m annealed silica feature in our
model, but we also note the size of the errorbars in the data at the top of the 12.6
$\mu$m feature do not preclude a more rounded shape like the feature belonging to
annealed silica.  The 12.6 $\mu$m complex of $\alpha$-quartz does evolve from
double-peaked at 295K to somewhat trapezoidal and eventually single-peaked at
progressively higher temperatures of 505K, 620K, 785K, and 825K; the feature is
single-peaked and centered at 12.7 $\mum$ for $\beta$-quartz.  As opposed to
annealed silica, $\alpha$- and $\beta$-quartz in the CDE shape distribution do not
fit the 20 $\mu$m feature, and both quartzes in the CDE2 shape distribution do not
fit the 9 $\mu$m feature (see discussions in sections 4.1 and 4.2 on ROXs 42C and ZZ
Tau, respectively).  Further, models with annealed silica fit the 16 $\mu$m features
of 1RXS J161410.6-230542 and ROXs 42C noticeably better than does quartz in either the 
CDE or CDE2 shape distributions.

\subsection{IRS 49}

IRS 49, also known as WLY 2-49, has been searched for multiplicity but has not been
found to be multiple \citep{bars03,ratz05}.  \citet{ratz05} established upper limits
on the relative brightness at K band of any unseen companion around IRS 49 of 0.04
for 0.15\arcsec\, separation and 0.02 for 0.50\arcsec\, separation.  \citet{gatti06}
report significant accretion of $\mdot$ = $1.3\times10^{-8}$ $\msunyr$ from
Paschen-$\beta$ and Brackett-$\gamma$ emission.

IRS 49 is in the Ophiuchus star-forming region; it has a sizeable extinction of
A$_{V}$ of 10.7 (Table 1).  How extinction at visible wavelengths translates to
extinction at mid-infrared wavelengths is still uncertain (see discussion by
\citet{d03} on values of A$_{V}/\tau_{9.7}$ determined from various studies); for
this reason, we present fits to the spectrum of IRS 49 dereddened assuming
A$_{V}/\tau_{9.7}$=18 (Figure 9, top, brown points and solid black line; $\chi^{2}$
per degree of freedom of 5.5) and assuming A$_{V}/\tau_{9.7}$=25 (Figure 9, top,
orange points and solid violet line; $\chi^{2}$ per degree of freedom of 6.3).  As
discussed by \citet{sarg06} with regard to the dust model of V410 Anon 13, greater
extinction correction of at mid-infrared wavelengths results in spectral profiles
that look more like the interstellar medium profile; namely, more like amorphous
silicates.  The effect of extinction correction on the silica features is most
important for the 9 and 20 $\mu$m silica features, but, as can be seen from 
Table 2 and Figure 7, $\chi^{2}$ per degree of freedom is lowest for both 
extinction corrections using models with annealed silica.

\subsection{T51}

Lastly, we consider T51.  Also known as Sz 41, this object was found to be double by
\citet{rz93}.  Another star, named Sz 41C, had also been thought to be associated
with Sz 41, but was shown to be a background star by \citet{walt92}.  This is noted
by \citet{corr06}, whose observations show the Sz 41 binary to be separated by
1.974\arcsec.  \citet{guen07} assign EWH$\alpha$ of -2.0\Angstrom, i.e., 
absorption.  Spectroastrometry conducted by \citet{tak03} determine that the 
H$\alpha$ profiles from the two components of the Sz 41 binary are roughly 
equivalent, suggesting there is negligible accretion in the disk(s) of the Sz 41 
system.

We fit models with different silicas to T51.  The spectrum of T51 includes a very
prominent peak at 9 $\mu$m, a triangular 20 $\mu$m feature, a modest 12.6 $\mu$m
feature, and a slight 16 $\mu$m feature.  Once again, the model with annealed silica
fit the dust excess spectrum the best (see Figure 9, bottom; $\chi^{2}$ per degree
of freedom of 3.6).  However, we note that of all objects whose spectra were
analyzed in this study, T51 provides the weakest tests discriminating between the
various versions of silica (see Figure 7).  The models using obsidian and amorphous
SiO$_2$ are not much worse than the model using annealed silica.

\section{Discussion}

\subsection{Cristobalite Dominance}

All the models (see Table 2) indicated the presence of both amorphous and
crystalline silicates and grains grown to larger sizes as well as silica, suggesting
production of silica accompanies both dust processing and grain growth.  Models
using the opacity of annealed silica provided the best fit of all models for all
five objects whose spectra we analyzed.  From our discussion of the KBr effect in
Section 3.2, we believe the opacity of the same silica grains in vacuum instead of
KBr would be very similar; any difference could be compensated by assuming more
extreme ellipsoidal shapes (as in CDE).  We conclude that the silica polymorph
giving rise to the silica features we see in the spectra of ROXs 42C, ZZ Tau, 1RXS 
J161410.6-230542, IRS 49, and T51 is predominantly cristobalite because cristobalite
was the dominant polymorph of silica present in the annealed silica sample of
\citet{fab00}, although an admixture of tridymite is possible.

\subsection{Silica in the Solar System}

\citet{dodd81} reports that tridymite and cristobalite are found in chondritic
meteorites \citep[see also][]{binns67}; more specifically, they were found in
enstatite chondritic meteorites, which are characterized by calcium-aluminum
inclusions \citep[CAIs;][]{guan00}.  \citet{dodd81} also summarizes findings 
on eucrite meteorites.  Eucrites are achondrites, which means they lack chondrules.  
Chondrules are spheroidal inclusions in chondritic meteorites, are submillimeter 
to centimeter in diameter, and have the appearance of having formed as molten 
drops \citep{nort02}.  The findings show the volume percentage of silica (as 
quartz, tridymite, and cristobalite polymorphs) in these eucrites to be between 
0 and 4 \%.  In the review by \citet{brjo98}, it is mentioned that less 
than 2\% of chondrules are silica-bearing chondrules, chondrules that are up 
to 40\% by volume silica, and they have been observed in a number of ordinary 
chondrites.

Silica has not been widely reported as detected in the infrared spectra of comet 
comae \citep{hanner03,habr04}.  Neither Comet Kohoutek \citep{merr74} nor Comet 
Hale-Bopp \citep{har02} showed any hint of the 9 and 12.6 $\mu$m silica features.  
{\it Spitzer} IRS spectra of the comet nucleus ejecta from the Deep Impact 
mission to Comet Tempel 1 also show no hint of silica features, and models of 
emission from this dusty eject do not require silica for a good fit \citep{lisse07}.  
\citet{bouw01} reported silica present based on the spectrum of Comet Halley;
however, the data is not of high quality, being coarse and noisy, so the detection
is tentative at best.  As discussed by \citet{har02}, magnesium-rich crystalline 
silicates absorb stellar radiation poorly over visible and near-infrared 
wavelengths, where the Sun emits a large fraction of its radiation.  \citet{har02} 
argue that Mg-rich pyroxene should have optical constants very similar to those 
of amorphous Mg-rich pyroxene given by \citet{dor95}.  The imaginary part of the 
index of refraction is as low as $\simali$ 0.0003 at 0.5 $\mum$ wavelength for 
amorphous Mg$_{0.95}$Fe$_{0.05}$SiO$_3$, so we follow \citet{har02} and take this 
value of k for Mg-rich crystalline pyroxene, which is found in comets 
\citep[see][]{har02}.  Optical properties for the various silica polymorphs at 
visible and near-infrared wavelengths are not readily available in the literature 
(macroscopic rocks of quartz are often transparent or translucent).  We estimate an 
upper limit on k for both ordinary and extraordinary rays of $\alpha$-quartz of 
10$^{-3}$ \citep[see relevant discussion by][]{palik85} and for amorphous SiO$_2$ 
of 10$^{-5}$ \citep[again see][]{palik85}.  This suggests that, at wavelengths 
over which the Sun emits most of its power, any silica in comets is either as 
poorly absorptive as Mg-rich pyroxenes, which are seen in comets, or is even less 
absorptive.  Thus, silica may be present in comets but may be at temperatures too 
low to emit enough to be seen in mid-infrared spectra of comet comae.

A better approach to searching for silica from comets is to analyze physical 
samples of comet dust.  {\it In situ} measurements of comet dust to date do 
not determine mineralogy, only the abundances of elements constituting comet 
dust.  Such measurements determined the rocky material in the coma of Comet 
Halley to have solar, or chondritic, abundances of the major rock-forming 
elements Mg, Si, Ca, and Fe \citep{hanner03}.  The STARDUST mission to Comet 
81P/Wild 2 returned one several-micron sized grain (``Ada'') composed of 
tridymite and fayalite \citep{zol06}.  \citet{mikou07} report a grain of 
crystalline silica that is either tridymite or cristobalite or both from 
Comet 81P/Wild 2.  It remains to be seen how much silica will be inferred 
for this comet based upon analysis of the returned samples of its dust.

As \citet{brad03} summarizes, some Interplanetary Dust Particles (IDPs) are 
believed to originate from comets.  \citet{rimc86} reported a 1.5 $\mum$ size 
grain of SiO$_2$ in the phenocryst of the IDP W7010*A2 constituting less 
than 5\% of the abundance of all minerals in the IDP.  This SiO$_2$ grain is 
of unknown polymorph \citep{rimc85}.  \citet{rimc86} also reported SiO$_2$ 
fragments of $<$ 10\% abundance in the same IDP.  These crystals are subhedral 
and $\simali$ 0.1 $\mum$ in diameter \citep{macriet87}.  However, \citet{brad03} 
does not mention silica as a significant mineral phase in IDPs.  \citet{grun01}, 
in the discussion of micrometeorite mineralogy and petrography, do not list 
silica as a major phase of micrometeorites, which are IDPs that have landed 
on Earth near the poles.

In a study of a large number of Kuiper Belt Objects (KBOs) using near-infrared 
spectra, \citet{bark08} found the spectra were well-fit by models assuming water 
ice and featureless spectral continuum from unknown material.  However, other 
compositions are inferred from near-infrared spectra of KBOs, such as methane 
and ethane ice, pyroxenes and olivines, and possible organic material 
\citep{emery07}.  It is unknown whether KBOs contain silica.

Cristobalite and tridymite commonly occur in terrestrial siliceous volcanic rocks,
such as in rapidly-cooled obsidian and rhyolite lava flows of San Juan district,
Colorado, USA, both as the lining of cavities and in the fine-grained groundmass
\citep[e.g.,][]{klhurl97}.

\subsection{Origin of Silica in Protoplanetary Disks}

Less than 5\% of the Si in the diffuse ISM is in crystalline silicates \citep{ld01},
and \citet{kemp04} did not detect silica in the interstellar medium 8-12.7 $\mu$m
absorption profile observed towards the Galactic Center.  The profile has no 9 or
12.3-12.6 $\mu$m features to indicate the presence of amorphous or crystalline
silica, and they very nicely fit the optical depth profile without silica.  This
optical depth profile was shown by \citet{sarg06} to be very similar to the 8-12.7
$\mu$m emissivities derived for CoKu Tau/4, DM Tau, and GM Aur, whose IRS spectra
were shown by \citet{daless05} and \citet{cal05} to be consistent with
protoplanetary disks with many-AU-sized almost empty inner holes (CoKu Tau/4 and DM
Tau) or optically thin inner regions and an almost empty gap (GM Aur).  The 10
$\mu$m silicate emission from these systems therefore originates from their outer
disk regions, where the dust is cooler and where dust processing is less likely to
occur.  Their 10 $\mu$m silicate emission profiles were very well fit with 
amorphous olivine and amorphous pyroxene and negligible silica, similar to the 
ISM absorption.  We infer the average starting mixture of grains in protoplanetary 
disks is like those of CoKu Tau/4, DM Tau, and GM Aur and has negligible amounts of 
silica.  Therefore, the silica we see must have formed in the disk.

In the creation of smoke from a laser-ablated enstatite grain, \citet{fab00} found
that the smallest smoke grains, 10-50 nm in size, were close to SiO$_{2}$ in
composition.  The smallest particles in a smoke created from a forsterite grain were
not pure SiO$_{2}$, but had Mg/Si ratios between 0.5 and 1.  The largest grains in
both smokes were close to the composition of their parent enstatite and forsterite
grains.  This could be an equilibrium process related to one in which enstatite
melts incongruently \citep{ba14} to forsterite plus a high temperature polymorph of
silica, either cristobalite or tridymite: 

\begin{eqnarray}
\mathrm{2MgSiO_{3}} & \rightleftharpoons & \mathrm{Mg_{2}SiO_{4}} + \mathrm{SiO_{2}}\\
\mathrm{(Enstatite)} &  & \mathrm{(Forsterite)} +
\mathrm{(Cristobalite~or~Tridymite)} \nonumber,
\end{eqnarray}

\noindent except that in this case amorphous pyroxene,
[Mg$_{1-x}$,Fe$_{x}$]SiO$_{3}$, replaces MgSiO$_{3}$, and amorphous olivine,
[Mg$_{1-x}$,Fe$_{x}$]$_{2}$SiO$_{4}$, replaces Mg$_{2}$SiO$_{4}$ \citep{bowsch35}. 
Here, x is probably around 0.5 -- we will call this reaction ``incongruent melting
of amorphous pyroxene''.  We invoke amorphous pyroxene as the starting
material in this reaction because amorphous pyroxene along with amorphous
olivine are hypothesized to be the main ingredients of the interstellar medium
\citep[ISM; see][]{kemp04} and the starting mixture of dust for protoplanetary disks
\citep[see discussion of unprocessed dust of transitional disks CoKu Tau/4, DM Tau,
and GM Aur by][]{sarg06}.  We also allow that incongruent melting of enstatite could
be happening, where the parent enstatite could itself be the daughter product of
prior processing of ``grandparent'' amorphous silicate.  Also, shocked enstatite
grains in meteorites are known to have formed cristobalite inclusions 
\citep{benzer02}.  Shock heating within the
protoplanetary disk \citep{hd02}; heating during disk-processing, flares, and
lightning \citep{pil98}; heating by intense light during the grains' removal from
the vicinity of the central star(s) by the T Tauri phase solar wind
\citep{depater01} or by an X-wind \citep{ssl96} all may cause melting of enstatite.
 Experiments by \citet{riet86}, \citet{hnd98}, \citet{hn98}, \citet{riet99}, and
\citet{riet02} also found silica grains (amorphous SiO$_2$ and tridymite) condensed
from vapors of silicate compositions.  This suggests silica can form in
protoplanetary disks as a smoke or condensate from parent silicate dust grains,
preferentially of low Mg/Si ratio like pyroxene.  However, silica is not expected
to condense from vapor of solar composition in chemical equilibrium \citep{gail04} 
in inner disk regions, 
suggesting the silica grains in the protoplanetary disks in our study did not arise
from chemical equilibrium condensation in the solar nebula close to the Sun.

\subsection{Silica as a Tracer of Dust processing}

After condensation of grains of the various silicates and silica, the experiments by
\citet{riet86}, \citet{hnd98}, \citet{hn98}, \citet{riet99}, \citet{fab00}, and
\citet{riet02} subjected the condensates to high temperatures to anneal them.  The
two experiments of annealing at the highest temperatures for the longest durations,
the heating of silica from iron-silicate smokes at 1300K for 120 hours by
\citet{hnd98} and of pure silica grains at 1220K for 5 hours by \citet{fab00}, gave
very similar opacity profiles.  \citet{fab00} determined that their annealed silica
was mostly cristobalite with some tridymite.  The similarity of the opacity spectra
indicates a similar composition for the \citet{hnd98} annealed silica.  The silica
grains remaining after annealing at lower temperature for lesser durations in the
other annealing experiments of \citet{fab00} remained tridymite and amorphous
SiO$_2$.  Tridymite forms at lower temperatures and does not have an appreciable 16
$\mum$ feature.  Therefore the 16 $\mu$m features of 1RXS J161410.6-230542 and ROXs 
42C, which were better fit with annealed silica (which has a 16 $\mu$m feature) 
than with other versions of silica, suggests the presence of at least some 
cristobalite in our silica exemplars.  Thus, we expect the silica grains in these 
protoplanetary disks were subjected to temperatures for durations of time like the 
most extreme annealing experiments of \citet{hnd98} and \citet{fab00}.

One possibility is that tridymite or cristobalite forms close to the central star,
where the stellar irradiation heats grains to the high temperatures required for
formation of tridymite or cristobalite, and then are transported to cooler disk
regions \citep{bm02}.  Our dust models have annealed silica (cristobalite/tridymite
admixture) at temperatures ranging from 153K to 1401K, corresponding to distances
from the star in the disk of a few AU down to the dust sublimation radius (sub-AU
range).  To explain the existence of tridymite or cristobalite in the cool outer
disk regions, however, the tridymite or cristobalite, once formed, would have to be
cooled sufficiently quickly so that their crystalline structure does not revert to
that of either of the lower-temperature polymorphs of silica, $\beta$- and
$\alpha$-quartz, and then transported to the outer disk.  It seems unlikely that any
of the proposed transport mechanisms could move these silica grains to a lower
temperature region of the disk rapidly enough to quench the cristobalite or 
tridymite structure in place.

If the tridymite or cristobalite do not form in the inner disk, this would suggest
transient heating events, like shocks \citep{hd02} or lightning \citep{pil98},
occuring {\it in situ} in the outer disk.  The shock model of \citet{dc02} provides
cooling similar to that required to explain chondrules - rates of 10-1000 K/hour -
which may be quite sufficient to allow silica grains to retain their
high-temperature polymorph crystalline structure.  Furthermore, some of the
\citet{dc02} shock models reached temperatures high enough that chondrule precursors
would be evaporated; this may allow formation of silica {\it via} condensation.  The
shock model of production of crystalline silicate grains has been invoked to explain
the presence of crystalline silicates in outflows from Asymptotic Giant Branch (AGB)
stars by \citet{edgar08}.  The lightning theory by \citet{pil98} also attempts to
explain chondrule formation and therefore meets similar requirements for cooling
rates.  The heating of pyroxene in some Interplanetary Dust Particles (IDPs) to 
temperatures above 1258K and their subsequent rapid cooling of $\simali$1000 K per 
hour \citep{ppfive} also suggest heating and rapid cooling in protoplanetary disks, 
as the chondrules do.  We note again the chondrules bearing silica by up to 40\% by 
volume and speculate there may be a connection between the mechanism that formed 
submicron silica grains in our five silica exemplars and the mechanism that gave 
rise to the silica-bearing chondrules.  An experiment like the one that produced the 
annealed silica whose opacity we use showed that such annealed silica cooled from 
1223K to 533K over a span of 4 hours 20 minutes with average cooling rates dropping 
from $\simali$2000 K per hour initially to $\simali$ 50 K per hour at the end (H.
Mutschke, private communication).  Our data indicate silica grains have experienced
high temperatures required to form cristobalite or tridymite, and cooled quickly
enough to quench them.

In the various proposed silica-production mechanisms, the silica forms from parent
amorphous silicate grains of pyroxene ([Mg,Fe]SiO$_{3}$) or olivine
([Mg,Fe]$_{2}$SiO$_{4}$) composition like what are inferred to exist in the ISM from
extinction profiles of objects in the Galactic Center \citep{kemp04} or from parent
crystalline grains of enstatite (MgSiO$_{3}$) or forsterite (Mg$_{2}$SiO$_{4}$). 
Other products arising from the production of silica (SiO$_{2}$) must be rich in the
leftover magnesium, iron, and/or oxygen.  This is seen, for example, in the
previously-discussed incongruent melting of enstatite or amorphous pyroxene,
which produces silica and also forsterite; the forsterite has the magnesium and
oxygen from the parent enstatite that the daughter silica lacks.  This would suggest
a correlation between silica abundance and silicates with higher stoichiometric
ratios of magnesium-to-silicon and oxygen-to-silicon like forsterite and the
olivines; such correlations should be sought in future dust composition studies 
of protoplanetary disks.

\section{Summary and Conclusions}

We have analyzed the dust composition of five T Tauri stars using their {\it
Spitzer} Space Telescope IRS spectra, with special regard to their silica dust. 
These five spectra show very prominent emission features at 9, 12.6, 20, and
sometimes 16 $\mu$m wavelength, which is characteristic of submicron size silica
dust.  We have constructed spectral models for each of the spectra that include
blackbodies at two temperatures, to represent inner and outer disk emission from the
optically thick disk midplane and blackbody grains, and Planck functions at those
two temperatures multiplied by scaled dust emissivities, to represent inner and
outer disk emission from dust in the optically thin disk atmosphere with strong
infrared resonances.  The best fit for a given set of dust opacities happens at the
pair of temperatures for which the global $\chi^{2}$ per degree of freedom is
minimum.  The opacities we use for non-silica dust species are either the same or
slightly improved over those used by \citet{sarg06}.

We conclude the following:

\begin{itemize}
\item The spectra rule out the highest pressure polymorph of silica, stishovite, and
we also fairly confidently rule out the intermediate pressure polymorph of silica,
coesite.
\item Amorphous versions of silica like amorphous SiO$_2$ \citep{hm97} and
silica-rich glass, obsidian \citep{koike89} are quite firmly ruled out to explain
the silica features in our five silica exemplars due to their inability to fit the
12.6 $\mu$m features seen in these spectra.  The amorphous SiO$_2$ features peak at
too short of a wavelength for this feature, are too wide, and are too weak with
respect to the 9 $\mu$m feature.
\item The opacity of annealed silica obtained by heating amorphous SiO$_2$ for 5
hours at 1220K \citep{fab00} provides the best fit to the spectra of all five of the
strong silica exemplars in this study.  \citet{fab00} report this annealed silica to
be mostly cristobalite but partly tridymite, so these silica polymorphs are the best
candidates to account for the silica features seen in the spectra of our five silica
exemplars.  This opacity is very similar to that reported by \citet{hnd98} for
silica annealed at 1300K for 120 hours.
\item The opacity of $\alpha$-quartz at intermediate temperatures (505K, 620K)
provided a good fit to the 12.6 $\mu$m feature of ROXs 42C, but the fits to features
at other wavelengths characteristic of silica (9, 16, 20 $\mu$m) were not as good as
the fits obtained from using annealed silica, indicating $\alpha$-quartz is not the
dominant silica in these disks.  $\beta$-quartz, however, can be fairly firmly ruled
out, as its feature at 12.7 $\mum$ cannot fit the 12.6 $\mum$ feature we observe in
our dust excess spectra, regardless of the assumed shape distribution of the
$\beta$-quartz grains.
\item Neither amorphous nor crystalline silica are seen in interstellar medium dust,
so the silica present in protoplanetary disks of the five silica exemplars analyzed
in this study must have formed in the disks.
\item The silica grains giving rise to emission from our five silica exemplars may
have formed in similar conditions to those that gave rise 
to Ada, a grain composed mostly of tridymite surrounded by fayalite, found in the
coma of Comet Wild/2 and the silica polymorphs found in primitive meteorites.
\item Tridymite and cristobalite are common products of annealing of amorphous dust
smokes and condensates in laboratory experiments (e.g, \citet{riet86},
\citet{riet99}, \citet{fab00}, \citet{riet02}) designed to mimic protoplanetary disk
conditions thought to produce crystalline silicates by annealing amorphous silicate
smokes.  Cristobalite, the major component of the annealed silica we use, is present
in silica grains annealed at temperatures above 1220K for durations more than 5
hours and has a 16 $\mu$m feature in its opacity profile that is needed to fit the
spectra of our silica exemplars.  Silica, as a tridymite or cristobalite polymorph,
might form in protoplanetary disks in manners similar to these laboratory
experiments.  Protoplanetary disk temperature and heating duration conditions giving
rise to the silica we see in our sample of protoplanetary disks must be more like
the most extreme conditions from the annealing experiments giving rise to
cristobalite.
\item The silica grains may form {\it via} the incongruent melting reaction of
amorphous pyroxene or the incongruent melting reaction of enstatite.
\item The tridymite or cristobalite grains, once formed, must be cooled quickly
enough to retain their crystalline structure and not revert to the lower temperature
polymorphs of silica, $\beta$- and $\alpha$-quartz.  Mechanisms like spiral shocks
and lightning devised to explain the rapidity with which chondrules cool may also
satisfy the rapidity with which tridymite or cristobalite must cool in order to
retain its crystalline structure.  This would support the idea of {\it in situ}
formation of the annealed silica.
\end{itemize}

\acknowledgments This work is based on observations made with 
the {\it Spitzer Space Telescope}, which is operated by 
the Jet Propulsion Laboratory, California Institute of Technology 
under NASA contract 1407.  Support for this work was provided by 
NASA through Contract Number 1257184 issued by JPL/Caltech and 
through the Spitzer Fellowship Program, under award 011 808-001, 
and JPL contract 960803 to Cornell University, 
and Cornell subcontracts 31419-5714 to the University of Rochester.  The
authors wish to thank Harald Mutschke for sharing the opacity in tabular 
form of the annealed silica presented by \citet{fab00} and for sharing cooling data
for the annealed silica.  A.L. acknowledges support from the Chandra theory program,
the Hubble theory programs, and the Spitzer theory programs.
ARB acknowledges NSF grants.  In addition, the authors thank our anonymous referee 
for a thoughtful review that greatly improved the paper.  SMART was developed by 
the IRS Team at Cornell University and is available through the Spitzer Science 
Center at Caltech.  This publication makes use of the Jena-St. Petersburg 
Database of Optical Constants \citep{hen99}.  The authors made use of the 
SIMBAD astronomical database and would like to thank those responsible 
for its upkeep.

\section*{Appendix}

For ZZ Tau, average flux densities for each DCE of each order of each nod were
computed, averaging over 5.18-6.76 $\mu$m wavelength for SL second order, 7.57-8.37
$\mu$m wavelength for SL bonus order, 7.57-11.81 $\mu$m wavelength for SL first
order, 14.15-17.97 $\mu$m wavelength for LL second order, 19.82-21.19 $\mu$m
wavelength for LL bonus order, and 22.02-29.99 $\mu$m wavelength for LL first 
order.  Each DCE of ZZ Tau was then scaled so that its average flux was the same 
average flux as the other DCEs of the same order.  For SL second order, all DCEs were 
scaled so that their average flux was the same as that of the second nod from the 
Campaign 4 AOR; for SL bonus order, the average flux to match was that of the first 
nod of the Campaign 4 AOR; for SL first order, also the first nod of the Campaign 4 
AOR; for LL second order, the second DCE of the second nod of the Campaign 29 AOR; 
for LL bonus order, the first DCE of the first nod of the Campaign 29 AOR; and for 
LL first order, the second DCE of the first nod of the Campaign 29 AOR.  In almost 
all cases, the average fluxes of these named DCEs were the highest of their given
order; in the case of LL second order, the average flux of the named DCE was only
1\% discrepant from the highest average flux of all LL second order DCE spectra. 
The range of scalars applied to each DCE of ZZ Tau is provided also in Table 3.

\clearpage

\begin{figure}[t] 
  \epsscale{1.0}
  \plotone{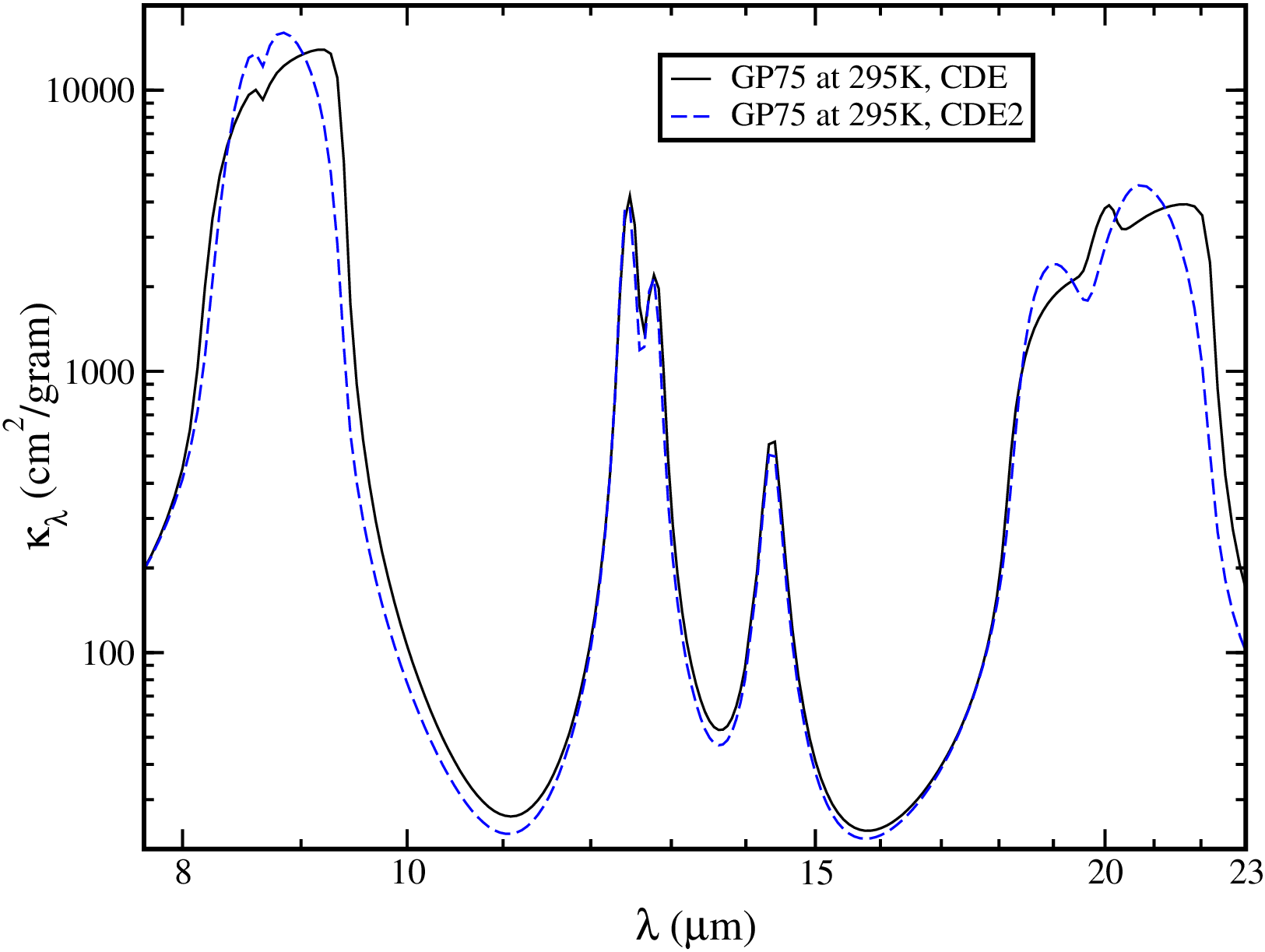}
  \caption{{\bf (Color on-line only)} Comparison of opacities of $\alpha$-quartz in the CDE and
CDE2 shape distributions using optical properties from \citet{gp75} for
$\alpha$-quartz at 295K.  Note the shifting of the long-wavelength sides
of the 9 and 20 $\mu$m features, but not of the weaker 12.6 and 14 $\mu$m 
features, to longer wavelengths.}
\end{figure}

\clearpage

\begin{figure}[t] 
  \epsscale{1.0}
  \plotone{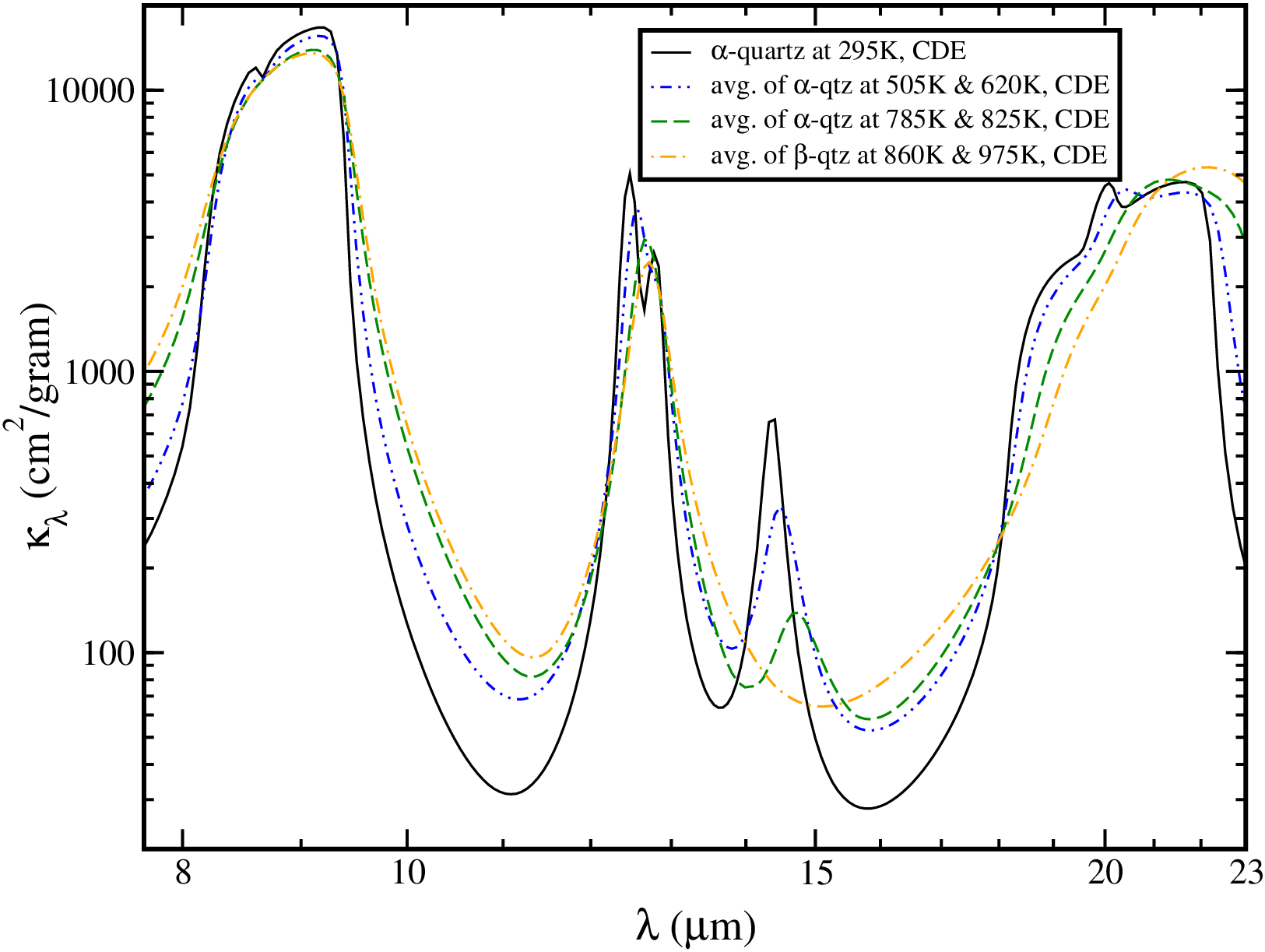}
  \caption{{\bf (Color on-line only)} Comparison between opacity profiles computed for grains in CDE 
shape distribution using optical constants of $\alpha$-quartz and $\beta$-quartz 
from Gervais and Piriou 1975.  The opacity profile of $\alpha$-quartz at 295K is 
plotted along with the average of the opacity profiles of $\alpha$-quartz at 505K
and 620K, 
the average of the opacity profiles of $\alpha$-quartz at 785K and 825K, and the 
average of the opacity profiles of $\beta$-quartz at 860K and 975K.}
\end{figure}

\clearpage

\begin{figure}[t] 
  \epsscale{1.0}
  \plotone{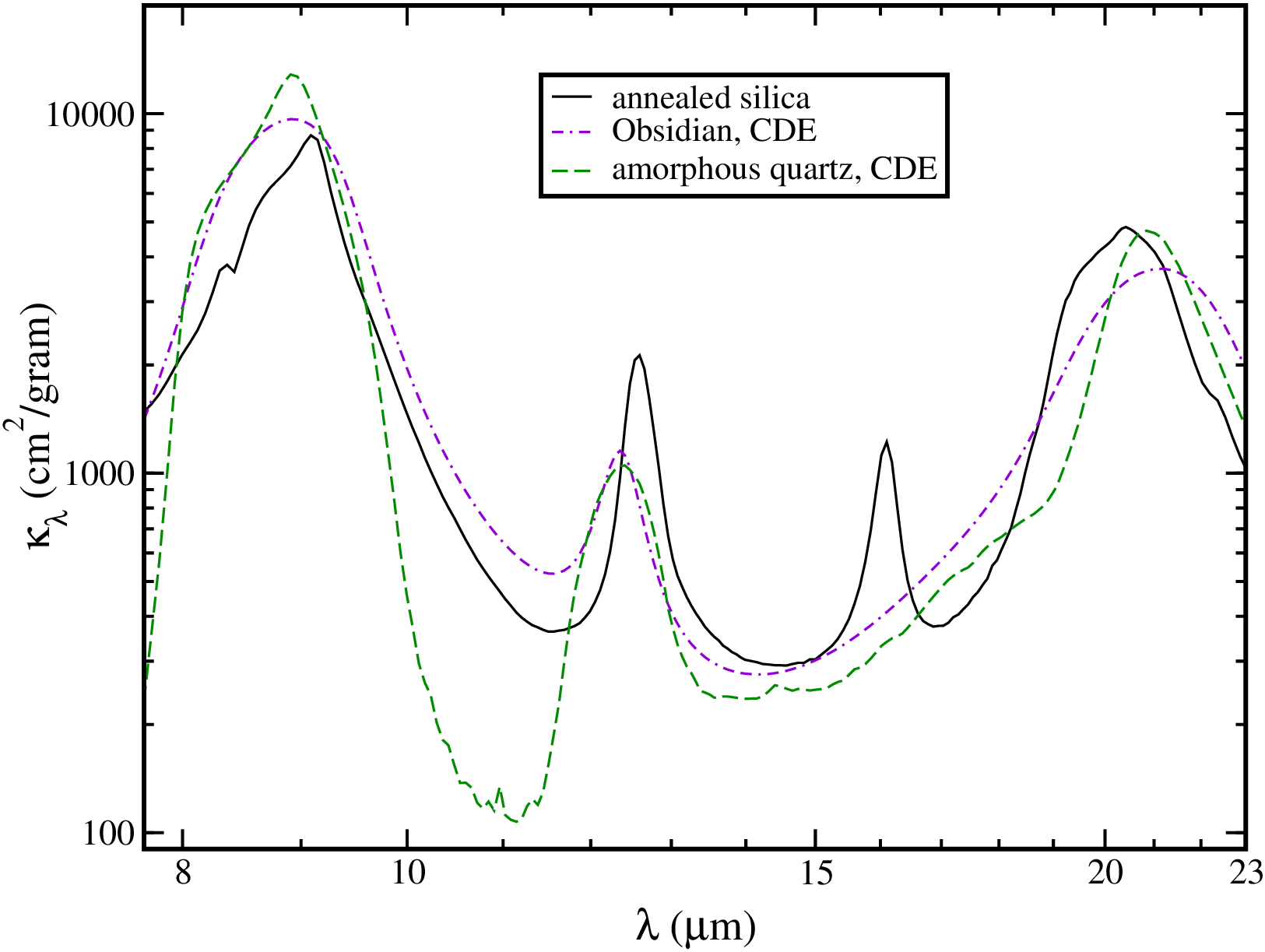}
  \caption{{\bf (Color on-line only)} Comparison of opacities of obsidian and amorphous SiO$_2$ in the CDE 
shape distribution and annealed silica.  Note the ratio of maximum opacity in the
12.3 $\mu$m feature
to the maximum opacity in the 9 $\mu$m feature for obsidian and amorphous SiO$_2$ is
lower than that of the 12.6 $\mu$m feature 
to the 9 $\mu$m feature for annealed silica.}
\end{figure}

\clearpage

\begin{figure}[t] 
  \epsscale{1.0}
  \plotone{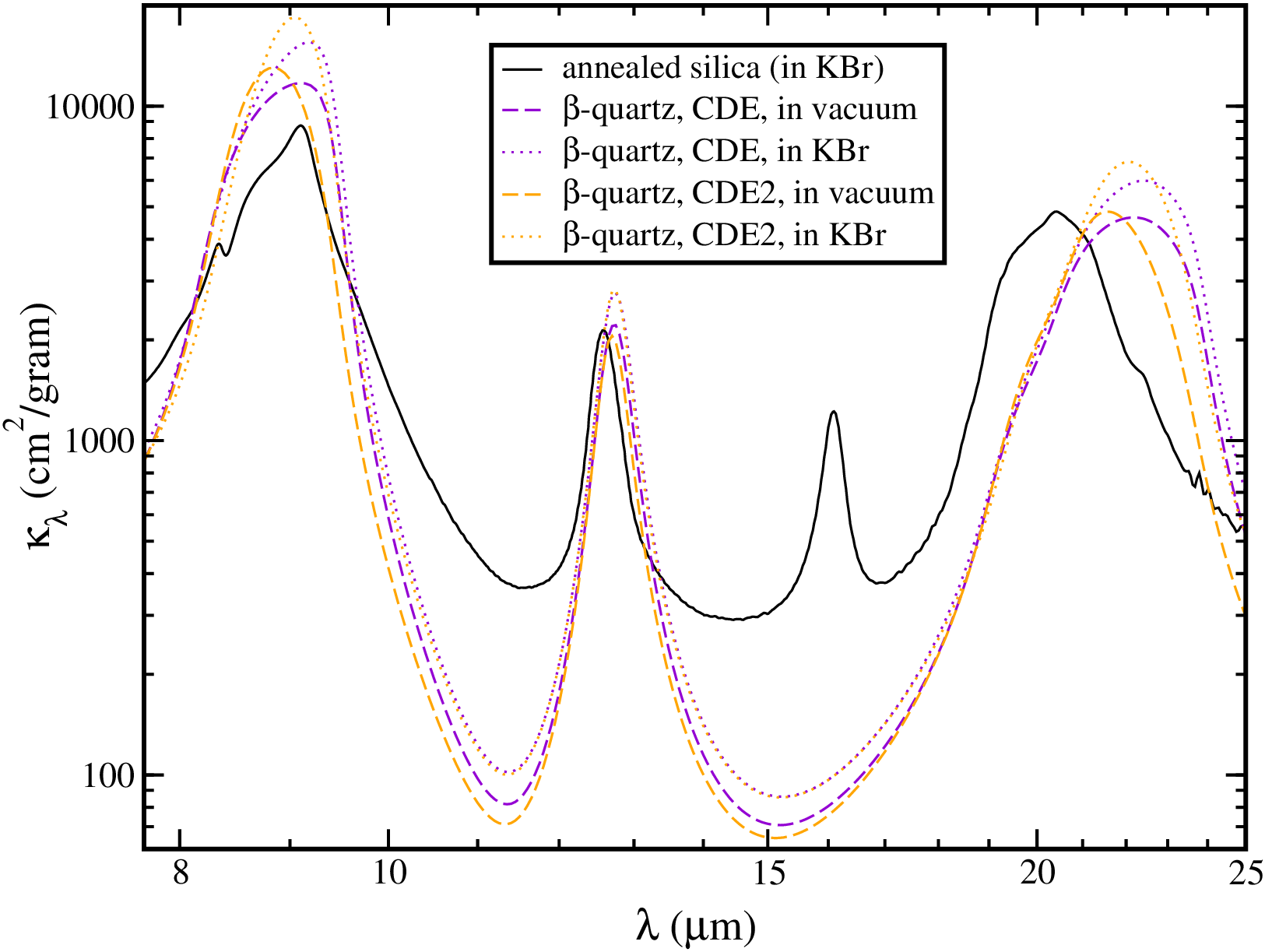}
  \caption{{\bf (Color on-line only)} KBr effect on $\beta$-quartz at 975K.  The peak wavelengths of the 9 $\mu$m 
features of $\beta$-quartz grains in both shape distribution in KBr and 
of annealed silica grains (in KBr) are very close.  The CDE2 shape distribution of 
$\beta$-quartz in vacuum gives a 9 $\mu$m peak $\sim$ 0.2 $\mu$m shortward of the
same grains 
in KBr, while the same peak for $\beta$-quartz in the CDE in vacuum is not much
different in 
central wavelength from the peaks of $\beta$-quartz in KBr.  The $\sim$ 20 
$\mu$m complexes of the various $\beta$-quartz opacity curves are similarly
affected; the weaker 
12.7 $\mu$m feature is hardly affected in terms of peak width and central wavelength
by shape 
or KBr/vacuum considerations at all.}
\end{figure}

\clearpage

\begin{figure}[t] 
  \epsscale{1.0}
  \plotone{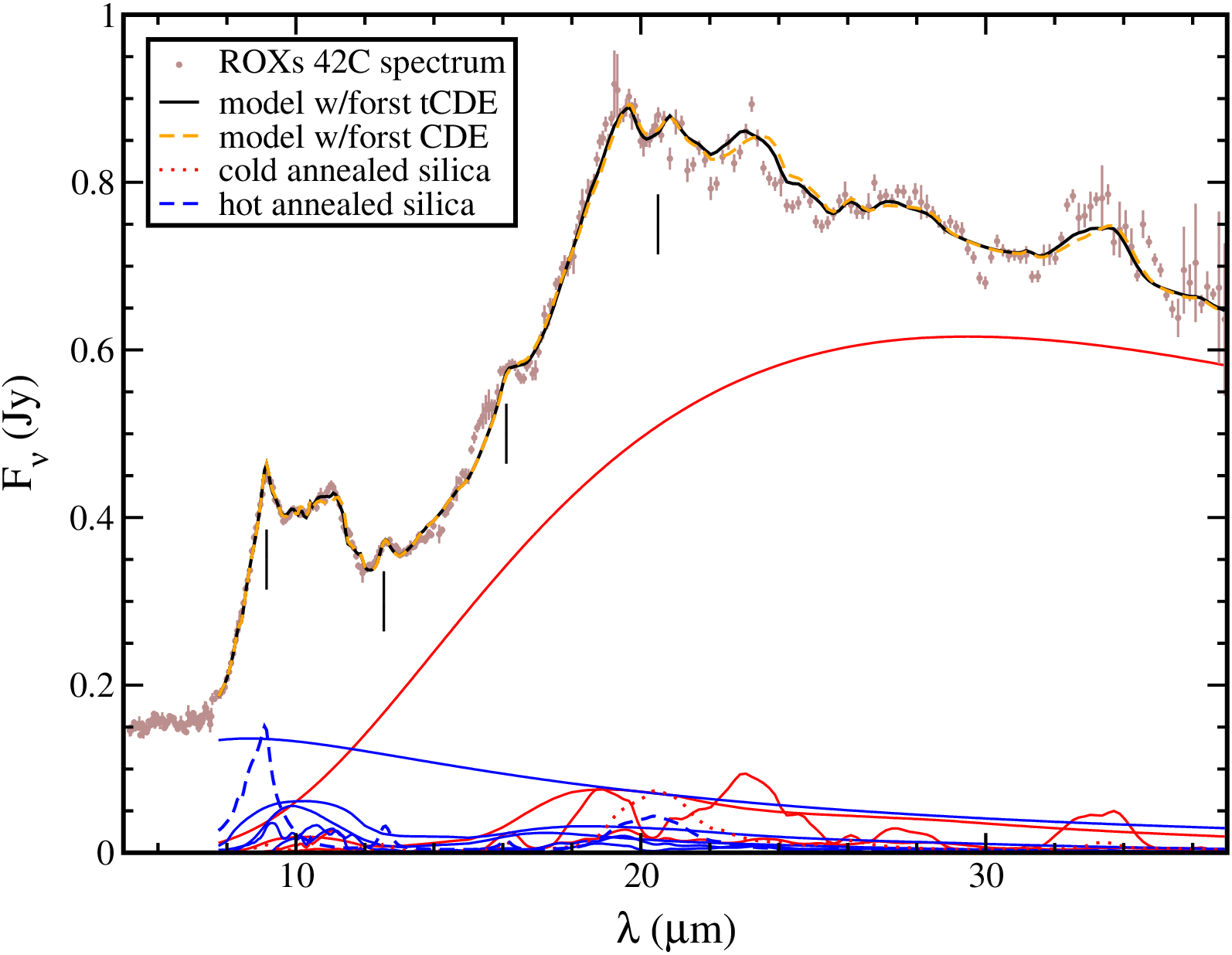}
  \caption{{\bf (Color on-line only)} Model fit to ROXs 42C using annealed silica, 
minimizing chi-squared from 6 to 37 $\mum$, comparing fit including forsterite in
the tCDE (black solid line; 
see text for description of tCDE) and CDE (orange dashed line) shape distributions. 
The model was constructed as described in Section 3 using parameters listed in Table
2.  All non-silica model components for the model using forsterite grains in the
tCDE shape distribution are plotted as solid lines at 
bottom of graph.  Model components from hot dust are plotted in blue, and those from
cold dust are plotted 
in red.  Silica components are plotted with a dashed line for the hot silica and a
dotted 
line for the cold silica.  Note the lack of the 6.2 $\mum$ PAH feature.}
\end{figure}

\clearpage

\begin{figure}[t] 
  \epsscale{1.0}
  \plotone{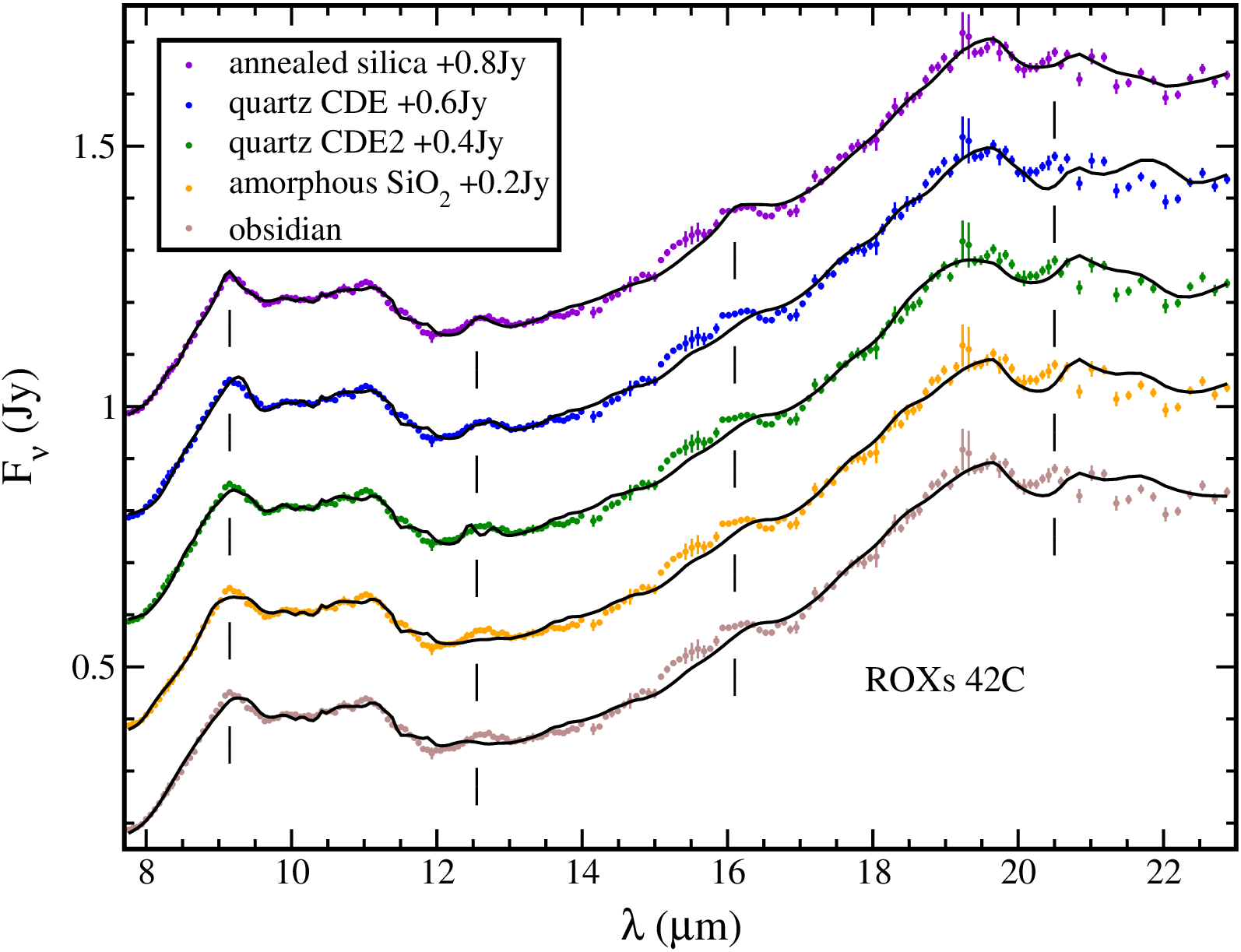}
  \caption{{\bf (Color on-line only)} Model fits to ROXs 42C.  The models are solid black lines, each model
using a different kind of silica and paired with the spectrum of ROXs 42C.  Each
model/spectrum pair is translated vertically by 0.2 Janskys.  The models use, from
bottom to top in the plot: obsidian, amorphous SiO$_2$, quartz in the CDE2 shape
distribution, quartz in the CDE shape distribution, and annealed silica.}
\end{figure}

\clearpage

\begin{figure}[t] 
  \epsscale{1.0}
  \plotone{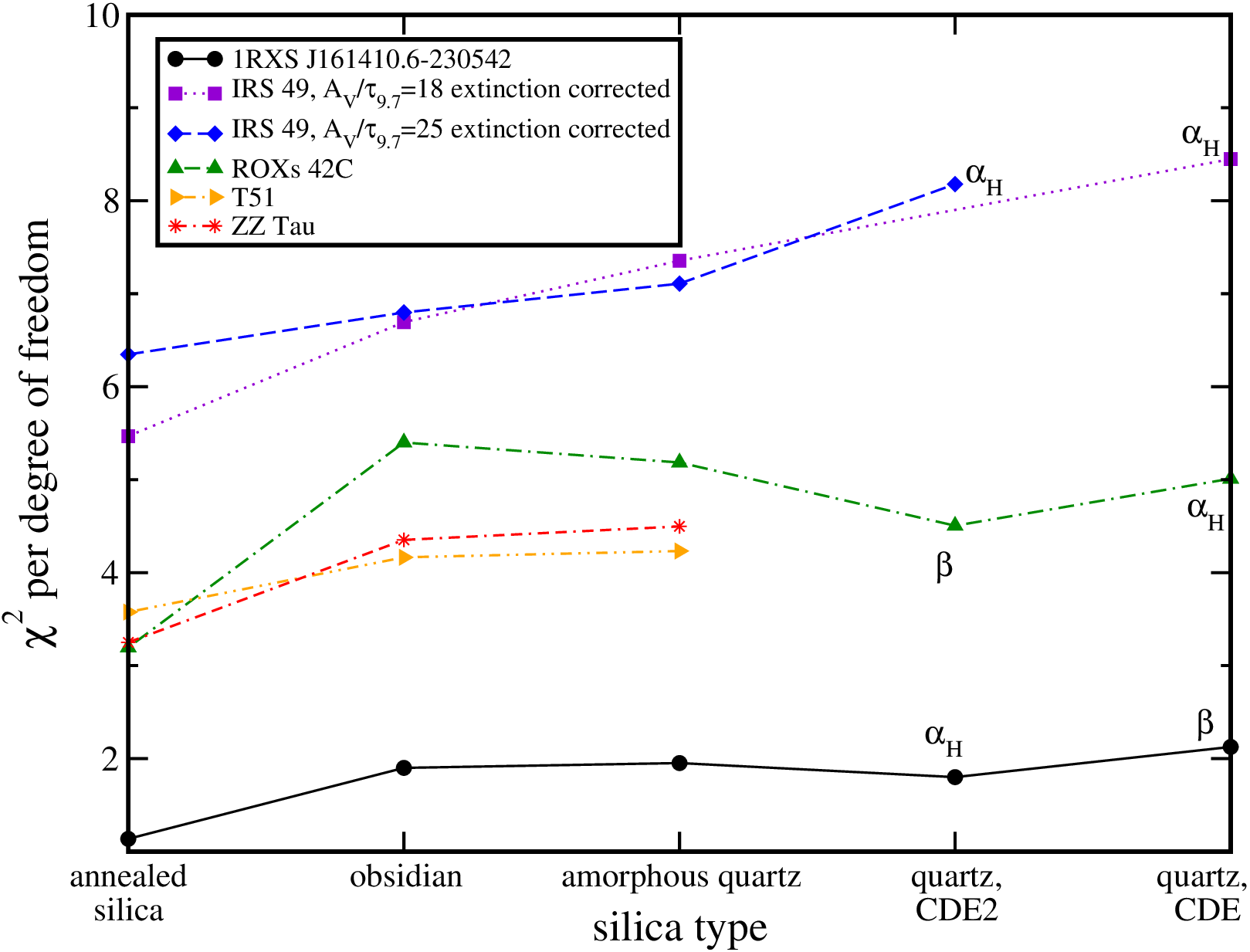}
  \caption{{\bf (Color on-line only)} Comparison of chi-squared per degree of freedom for dust emission models
using
different types of silica for the spectra of the five objects modeled, including the
dereddened
spectra obtain by two different computations of optical depth at 9.7 $\mum$ of the
material
in front of IRS 49.  The type of quartz - $\beta$-quartz and an average of
$\alpha$-quartz at 
785K and 825K - is denoted by $\beta$ and $\alpha_{H}$ respectively.}
\end{figure}

\clearpage

\begin{figure}[t] 
  \epsscale{0.82}
  \plotone{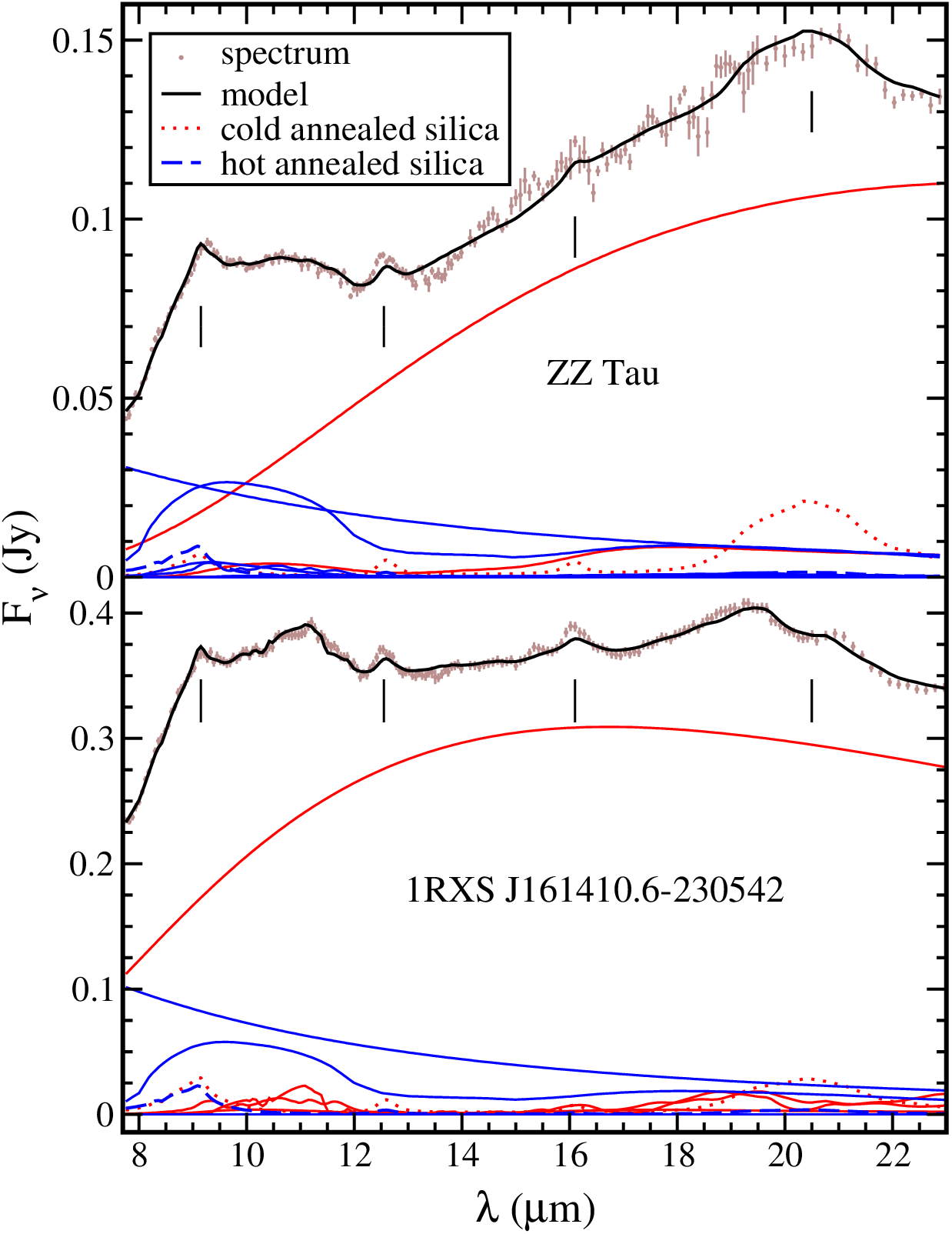}
  \caption{{\bf (Color on-line only)} Best fits to ZZ Tau (top) and 1RXS J161410.6-230542 (bottom) using annealed
silica.  Same color and linestyle scheme for model components as for
Figure 5.  The dominant non-continuum dust component in the 8-12 $\mum$ region of ZZ
Tau is large amorphous pyroxene.}
\end{figure}

\clearpage

\begin{figure}[t] 
  \epsscale{0.82}
  \plotone{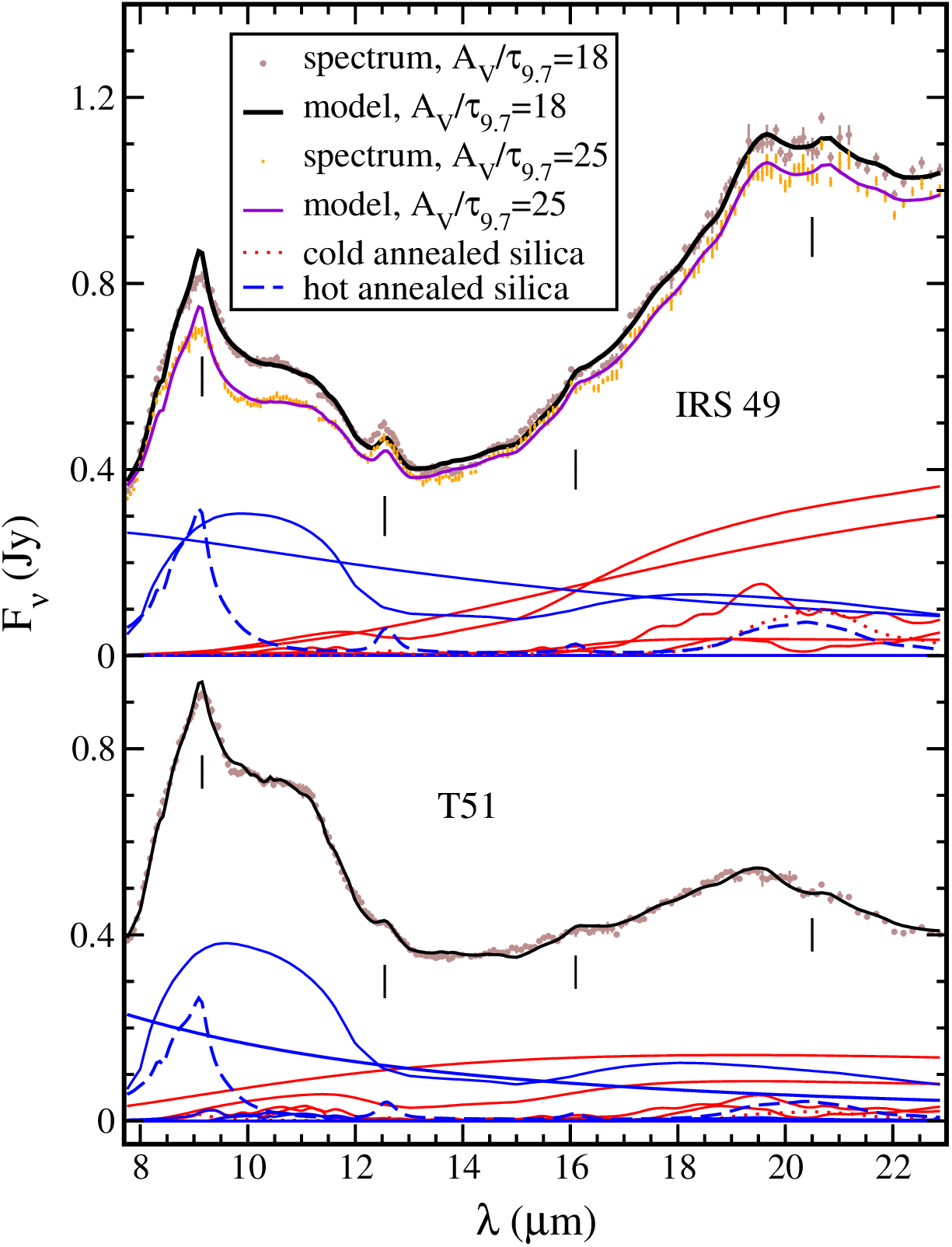}
  \caption{{\bf (Color on-line only)} Best fits to IRS 49 (top) and T51 (bottom) using annealed
silica.  Same color and linestyle scheme for model components as for Figure 5,
except for the orange points being the spectrum of IRS 49 extinction corrected
assuming A$_{V}/\tau_{9.7}$ = 25 instead of 18 
and the solid violet line running through the orange points being the model fit
using annealed silica to this differently-corrected data.}
\end{figure}

\clearpage

\begin{table}[h,t]
{\tiny
\caption[]{Observations and Stellar Characteristics \label{table1}}
\begin{tabular}{lcccccccccc}
\hline \hline
          
          & 
          & 
          & R.A.
          & Decl.
          & Spectral
          & 
          & 
          & 
          &
          & \\
          Object 
          & AOR 
          & \# in AOR 
          & (J2000)
          & (J2000)
          & Type 
          & A$_{V}$ 
          & T$_{eff}$
          & $\Omega_{star}$
          & dist.
          & Refs.\\
          (1)
          & (2)
          & (3)
          & (4) 
          & (5) 
          & (6) 
          & (7) 
          & (8)
          & (9)
          & (10)
          & (11)\\
\hline
1RXS J161410.6-230542 & 5206272 & \nodata & 16 14 11.08 & -23 05 36.2 & K0 & 1.48 &
4963 & 5.18 & 144 & 1,2,3\\
IRS 49 & 12698624 & 6 & 16 27 38.31 & -24 36 58.7 & K5.5 & 10.7 & \nodata & \nodata
& 140 & 4,5,6\\
ROXs 42C & 12676352 & 1 & 16 31 15.75 & -24 34 02.2 & K6 & 1.9 & 4205 & 8.54 & 140 &
6,7\\
T51 & 12696576 & 4 & 11 12 24.41 & -76 37 06.4 & K3.5 & 0 & \nodata & \nodata & 160
& 8,9\\
ZZ Tau & 3546880,16261376 & \nodata & 04 30 51.38 & +24 42 22.3 & M3 & 1.44 & 3470 &
4.90 & 140 & 10,11,12\\
\hline
\end{tabular}
\tablecomments{\footnotesize Col. (1): Object name.  Col. (2): {\it Spitzer}
Astronomical Observation 
                 Request number.  Col. (3): If multiple objects observed in AOR,
then specifies which
                 one observed.  If only one object in AOR, then no number specified.
 Col. (4): Right 
                 Ascension of object in J2000 coordinates.  Col (5): Declination of
object in J2000 
                 coordinates.  Col. (6): Spectral type of object.  Col. (7):
Extinction at V 
                 band in magnitudes.  Col. (8): If stellar photosphere emission was
subtracted from
                 the IRS spectrum, then this entry is the effective temperature of
the star which is 
                 the temperature of the blackbody representing stellar photosphere
emission normalized 
                 to the dereddened J-band flux density.  If stellar photosphere
emission not subtracted, 
                 then no effective temperature specified.  Col. (9): If stellar
photosphere emission 
                 subtracted, then solid angle in units of 10$^{-19}$ steradians of
blackbody representing 
                 stellar photosphere emission.  Obtained by dividing dereddened J
band flux density by 
                 Planck function evaluated at isophotal wavelength of J band and at
temperature specified 
                 in previous column; if no stellar photosphere emission subtracted,
then no solid angle
                 specified.  Col. (10): Assumed distance to object, in parsecs. 
Col. (11): References from 
                 which spectral type, A$_{V}$, and distance are obtained.  1 -
\citet{prei98}; 2 - \citet{pas07}; 3 - \citet{dezee99}; 4 -
\citet{wilk05}; 5 - \citet{mcc08}; 6 - \citet{bont01}; 7 -
\citet{ba92}; 8 - \citet{luh04}; 9 - \citet{luh08}; 10 -
\citet{hbc88}; 11 - \citet{fur06}; 12 - \citet{ken94}.  For 1RXS J161410.6-230542, 
T$_{eff}$ comes from \citet{pas07}; 
                 for ROXs 42C and ZZ Tau, T$_{eff}$ comes from \citet{kh95} based on
the spectral type from column 4.}
}
\end{table}

\clearpage

\begin{landscape}
\begin{deluxetable}{lcccccccccccc}
\tabletypesize{\scriptsize}
\tablewidth{680pt}
\tablecaption{Dust Model Parameters \label{table2}}
\tablehead{\colhead{} & \colhead{} & \colhead{} & \colhead{} & \colhead{} &
\colhead{small} & \colhead{small} & \colhead{large} & \colhead{large} & \colhead{} &
\colhead{} & \colhead{Total} & \colhead{}\\
\colhead{} & \colhead{Temp} & \colhead{} & \colhead{silica} & \colhead{} &
\colhead{Amorphous} & \colhead{Amorph.} & \colhead{Amorph.} & \colhead{Amorph.} &
\colhead{Crystalline} & \colhead{Crystalline} & \colhead{Dust} & \colhead{}\\
\colhead{Object} & \colhead{(K)} & \colhead{$\Omega_{BB}$} & \colhead{type} &
\colhead{Silica\tablenotemark{a}} & \colhead{Pyroxene\tablenotemark{b}} &
\colhead{Olivine\tablenotemark{c}} & \colhead{Pyroxene\tablenotemark{d}} &
\colhead{Olivine\tablenotemark{e}} & \colhead{Pyroxene\tablenotemark{f}} &
\colhead{forsterite\tablenotemark{g}} & \colhead{Mass} & \colhead{$\frac{\chi^2}{\rm
d.o.f.}$}\\
\colhead{(1)} & \colhead{(2)} & \colhead{(3)} & \colhead{(4)} & \colhead{(5)} &
\colhead{(6)} & \colhead{(7)} & \colhead{(8)} & \colhead{(9)} & \colhead{(10)} &
\colhead{(11)} & \colhead{(12)} & \colhead{(13)}}
\startdata
1RXS J161410.6 & 305 & 57.5 $\pm$ 0.8 & annSil & 35.7 $\pm$ 17.3 & 0 & 12.0 $\pm$ 16.3 & 0 & 0 & 26.1 $\pm$ 14.5 & 26.3 $\pm$ 14.2 & 0.815 & 1.1\\
-230542 & 1401 & 0.329 $\pm$ 0.020 & ``'' & 5.6 $\pm$ 3.2 & 0 & 0 & 94.4 $\pm$ 16.0 & 0 & 0 & 0 & 0.0474 & \nodata\\
IRS 49\tablenotemark{h} & 153 & 540 $\pm$ 34 & annSil & 6.2 $\pm$ 2.5 & 0 & 7.8 $\pm$ 3.6 & 0 & 70.6 $\pm$ 6.4 & 12.2 $\pm$ 2.6 & 3.2 $\pm$ 2.5 & 176 & 5.5\\
\nodata & 751 & 3.36 $\pm$ 0.11 & ``'' & 14.0 $\pm$ 1.5 & 0 & 0 & 86.0 $\pm$ 4.5 & 0 & 0 & 0 & 0.904 & \nodata\\
IRS 49\tablenotemark{i} & 160 & 404 $\pm$ 26 & annSil & 6.6 $\pm$ 2.3 & 0 & 0 & 0 & 79.7 $\pm$ 6.2 & 10.1 $\pm$ 2.3 & 3.6 $\pm$ 2.2 & 150 & 6.3\\
\nodata & 796 & 3.02 $\pm$ 0.09 & ``'' & 16.2 $\pm$ 1.9 & 0 & 0 & 83.8 $\pm$ 5.2 & 0 & 0 & 0 & 0.569 & \nodata\\
ROXs 42C & 191 & 429 $\pm$ 9 & annSil & 20.0 $\pm$ 5.4 & 11.6 $\pm$ 7.7 & 32.1 $\pm$ 7.4 & 0 & 0 & 14.6 $\pm$ 5.1 & 21.8 $\pm$ 5.5 & 25.3 & 3.2\\
\nodata & 1051 & 0.662 $\pm$ 0.046 & ``'' & 13.8 $\pm$ 2.6 & 5.4 $\pm$ 3.8 & 16.5 $\pm$ 4.5 & 57.1 $\pm$ 6.8 & 0 & 5.8 $\pm$ 3.3 & 1.5 $\pm$ 3.2 & 0.172 & \nodata\\
ROXs 42C & 172 & 613 $\pm$ 16 & $\alpha$qCDElt & 9.5 $\pm$ 6.4 & 0 & 37.7 $\pm$ 10.1 & 0 & 0 & 42.9 $\pm$ 10.5 & 9.9 $\pm$ 5.9 & 32.9 & 5.0\\
\nodata & 601 & 4.18 $\pm$ 0.15 & $\alpha$qCDEht & 18.4 $\pm$ 3.8 & 0 & 69.1 $\pm$ 11.4 & 0.1 $\pm$ 9.5 & 0 & 0 & 12.3 $\pm$ 4.4 & 0.393 & \nodata\\
ROXs 42C & 191 & 421 $\pm$ 9 & $\alpha$qCDE2lt & 15.1 $\pm$ 4.4 & 0 & 15.1 $\pm$ 5.5 & 0 & 39.5 $\pm$ 6.8 & 16.3 $\pm$ 4.2 & 14.1 $\pm$ 4.0 & 30.3 & 4.5\\
\nodata & 1151 & 0.707 $\pm$ 0.039 & $\beta$qCDE2 & 3.9 $\pm$ 2.3 & 70.3 $\pm$ 13.9 & 0 & 0 & 0 & 19.4 $\pm$ 5.9 & 6.4 $\pm$ 5.1 & 0.0900 & \nodata\\
ROXs 42C & 191 & 400 $\pm$ 9 & amsil & 17.9 $\pm$ 5.1 & 0 & 33.8 $\pm$ 7.0 & 0 & 0 & 29.7 $\pm$ 6.0 & 18.6 $\pm$ 4.8 & 27.4 & 5.2\\
\nodata & 751 & 1.86 $\pm$ 0.09 & ``'' & 10.3 $\pm$ 3.3 & 59.0 $\pm$ 12.4 & 17.4 $\pm$ 7.8 & 0 & 0 & 8.2 $\pm$ 5.8 & 5.2 $\pm$ 5.3 & 0.193 & \nodata\\
ROXs 42C & 210 & 275 $\pm$ 6 & obsid & 25.7 $\pm$ 4.1 & 21.8 $\pm$ 4.9 & 22.5 $\pm$ 4.5 & 0 & 0 & 15.7 $\pm$ 3.4 & 14.3 $\pm$ 3.2 & 24.8 & 5.4\\
\nodata & 1401 & 0.376 $\pm$ 0.028 & ``'' & 25.1 $\pm$ 8.1 & 0 & 62.8 $\pm$ 18.1 & 0 & 0 & 9.9 $\pm$ 8.1 & 2.3 $\pm$ 7.8 & 0.0427 & \nodata\\
T51 & 267 & 39.2 $\pm$ 1.7 & annSil & 4.8 $\pm$ 3.6 & 0 & 0 & 0 & 70.9 $\pm$ 9.3 & 15.2 $\pm$ 3.9 & 9.2 $\pm$ 3.6 & 6.14 & 3.6\\
\nodata & 1351 & 0.792 $\pm$ 0.037 & ``'' & 9.1 $\pm$ 1.1 & 0 & 0 & 88.2 $\pm$ 4.0 & 0 & 1.8 $\pm$ 1.5 & 1.0 $\pm$ 1.4 & 0.356 & \nodata\\
\tablebreak
ZZ Tau & 210 & 62.9 $\pm$ 1.5 & annSil & 45.8 $\pm$ 23.8 & 0 & 54.2 $\pm$ 27.1 & 0 & 0 & 0 & 0 & 1.46 & 3.3\\
\nodata & 1251 & 0.123 $\pm$ 0.007 & ``'' & 4.2 $\pm$ 1.9 & 5.5 $\pm$ 3.2 & 0 & 85.5 $\pm$ 8.2 & 0 & 4.2 $\pm$ 2.9 & 0.5 $\pm$ 2.8 & 0.0290 & \nodata\\
\enddata
\tablecomments{\footnotesize Col. (1): Object name.  Col. (2): One of two dust model
temperatures (Kelvin).  Col. (3): Solid angle, 
                 $\Omega_{BB}$, of blackbody of temperature specified in Col. (2)
representing 
                 continuum emission, expressed in units of $10^{-17}$ 
                 steradians.  Col. (4): Type of silica used in model.  Cols.
(5)-(11): mass percentages of all dust mass at temperature
specified in Col. (2).  One dust model is completely specified by
two adjacent rows - the row following the object$'$s 
                 name and the row beneath that one.  Col. (12): Total dust mass at
one temperature in $10^{-4}$ lunar masses, 
                 computed assuming distances to each object
                 as listed in Table 1.  Col. 
                 (13): $\chi^{2}$ per degree of freedom, determined over 
                 7.7\,$<$\,$\lambda$\,$<$\,23$\mum$.}
\tablenotetext{a}{\footnotesize Optical properties and opacities for various types
of silica from 
                  references given in text.  In the table, ``annSil'' means
                  annealed silica, ``$\alpha$qCDElt'' means an average of the
                  CDE opacity curves obtained for $\alpha$-quartz at 505K and 620K
from \citet{gp75};
                  ``$\alpha$qCDE2lt'' means the same, but for CDE2;
``$\alpha$qCDEht'' means an average of CDE opacity curves for
$\alpha$-quartz at 785K and 825K from 
                  \citet{gp75}; ``$\beta$qCDE2'' means the same, but for CDE2;
``obsid'' means obsidian; and ``amsil'' means amorphous SiO$_2$.}
\tablenotetext{b}{\footnotesize Optical constants for amorphous 
                 pyroxene Mg$_{0.7}$Fe$_{0.3}$SiO$_3$ 
                 from \citet{dor95}, 
                 assuming CDE2 \citep{fab01}}
\tablenotetext{c}{\footnotesize Optical constants for amorphous 
                 olivine MgFeSiO$_4$ from \citet{dor95}, 
                 assuming CDE2}
\tablenotetext{d}{\footnotesize Optical constants for amorphous 
                 pyroxene Mg$_{0.7}$Fe$_{0.3}$SiO$_3$ 
                 from \citet{dor95}, 
                 using the Bruggeman EMT and Mie theory
                 (Bohren \& Huffman 1983) 
                 with a volume fraction of vacuum of 
                 $f$\,=\,0.6 for porous spherical grains 
                 of radius 5$\mum$}
\tablenotetext{e}{\footnotesize Optical constants for 
                  amorphous olivine 
                  MgFeSiO$_4$ from \citet{dor95}, 
                  using the Bruggeman EMT and Mie theory
                 (Bohren \& Huffman 1983) 
                 with a volume fraction of vacuum of 
                 $f$\,=\,0.6 for porous spherical grains 
                 of radius 5$\mum$}
\tablenotetext{f}{\footnotesize Opacities for crystalline pyroxene 
                  Mg$_{0.9}$Fe$_{0.1}$SiO$_3$ from \citet{chi02}}
\tablenotetext{g}{\footnotesize Optical constants for 
                  3 crystallographic axes of forsterite, 
                  Mg$_{2}$SiO$_4$, 
                  from \citet{sog06}, assuming tCDE (see discussion in Section 3.3
of this paper) 
                  shape distribution.}
\tablenotetext{h}{\footnotesize Extinction corrected using 
                  A$_{V}/\tau_{9.7}$=18; $\tau_{9.7}$=0.6}
\tablenotetext{i}{\footnotesize Extinction corrected using 
                  A$_{V}/\tau_{9.7}$=25; $\tau_{9.7}$=0.4}
\end{deluxetable}
\end{landscape}

\clearpage

\begin{table}[h,t]
{\tiny
\caption[]{Mispointing Corrections \label{table3}}
\begin{tabular}{lcccccccc}
\hline \hline
          
          & Correction to 
          & `` ''
          & `` ''
          & `` ''
          & `` ''
          & `` ''
          & `` ''
          & `` ''\\
          Object 
          & SL2nod1 
          & SL2nod2 
          & SL1nod1 
          & SL1nod2 
          & LL2nod1 
          & LL2nod2 
          & LL1nod1 
          & LL1nod2\\
          (1) 
          & (2) 
          & (3) 
          & (4)
          & (5) 
          & (6) 
          & (7) 
          & (8) 
          & (9)\\
\hline
1RXS J161410.6-230542 & 1.03 & 1.03 & 1.03 & 1.03 & 1.00 & 1.00 & 1.00 & 1.00 \\
IRS 49 & 1.00 & 1.04 & 1.00 & 1.04 & 1.00 & 1.04 & 1.00 & 1.01 \\
ROXs 42C & 1.00 & 1.00 & 1.00 & 1.00 & 1.00 & 1.02 & 1.00 & 1.00 \\
T51 & 1.00 & 1.00 & 1.00 & 1.00 & 1.00 & 1.01 & 1.00 & 1.01 \\
ZZ Tau & 1.00-1.10 & 1.00-1.10 & 1.00-1.14 & 1.02-1.07 & 0.99-1.09 & 1.00-1.09 &
1.00-1.08 & 1.02-1.08 \\
\hline
\end{tabular}
\tablecomments{\footnotesize Col. (1): Object name.  Cols. (2)-(9): 
                 Multiplicative scalars applied to the spectrum of one order of one
nod for a given 
                 object to match the flux density of the other nod as described in
the text.  SL2 
                 means Short-Low order 2, SL1 is Short-Low order 1, LL2 is Long-Low
order 2, LL1 is 
                 Long-Low order 1.  For ZZ Tau, the range of scalars used for each
order of each nod,
                 derived as described in the text, is provided.}
}
\end{table}


\begin{thebibliography}{}

\bibitem[Alexander et al.(2007)]{ppfive} Alexander, C.~M.~O., 
Boss, A.~P., Keller, L.~P., Nuth, J.~A., 
\& Weinberger, A.\ 2007, Protostars and Planets V, 801
\bibitem[Armus et al.(2007)]{armus07} Armus, L., et al.\ 2007, 
\apj, 656, 148
\bibitem[Barkume et al.(2008)]{bark08} Barkume, K.~M., Brown, 
M.~E., \& Schaller, E.~L.\ 2008, \aj, 135, 55
\bibitem[Barsony et al.(2003)]{bars03} Barsony, M., Koresko, 
C., \& Matthews, K.\ 2003, \apj, 591, 1064
\bibitem[Benzerara et al.(2002)]{benzer02} Benzerara, K., et al.\ 2002, American
Mineralogist, 87, 1250
\bibitem[Binns(1967)]{binns67} Binns, R.~A.\ 1967, Am.\ Mineral.\ , 52, 1549
\bibitem[Bockel{\'e}e-Morvan et al.(2002)]{bm02} Bockel{\'e}e-Morvan, D., 
Gautier, D., Hersant, F., Hur{\'e}, J.-M., \& Robert, F.\ 2002, \aap, 384, 1107
\bibitem[Bohren \& Huffman(1983)]{bh83} Bohren, C.~F., \& 
        Huffman, D.~R.\ 1983, Absorption and Scattering of 
        Light by Small Particles, New York: Wiley
\bibitem[Bontemps et 
al.(2001)]{bont01} Bontemps, S., et al.\ 2001, \aap, 372, 173
\bibitem[Bouvier 
\& Appenzeller(1992)]{ba92} Bouvier, J., \& Appenzeller, I.\ 1992, \aaps, 92, 481
\bibitem[Bouwman et al.(2001)]{bouw01} Bouwman, J., Meeus, G., 
        de Koter, A., Hony, S., Dominik, C., \& Waters, L.~B.~F.~M.\  
        2001, \aap, 375, 950
\bibitem[Bowen \& Anderson(1914)]{ba14} Bowen, N.~L. \& Anderson, O.\ 1914, Am. J.
Sci, Series 4, 37, 487
\bibitem[Bowen \& Schairer(1935)]{bowsch35} Bowen, N.~L. \& Schairer, J.~F.\ 1935,
Am. J. Sci, Series 5, 29, 151
\bibitem[Bradley(2003)]{brad03} Bradley, J.\ 2003, LNP 
        Vol.~609: Astromineralogy, 609, 217
\bibitem[Brearley \& Jones(1998)]{brjo98} Brearley, A.~J. \& Jones, R.~H.\ 
(1998) Chondritic meteorites. In {\it Planetary Materials}, Reviews in 
Mineralogy (ed. J.~J. Papike). Mineralogical Society of America, 
Washington, D.~C., vol. 36, chap. 3, pp 3-13 and 3-37.
\bibitem[Calvet et al.(1992)]{cal92} Calvet, N., Magris, G.~C., 
         Patino, A., \& D'Alessio, P.\ 1992, 
         Rev. Mex. de Astro. Astrof., 24, 27
\bibitem[Calvet et al.(2005)]{cal05} Calvet, N., et al.\ 
2005, \apjl, 630, L185
\bibitem[Chen et al.(2006)]{chen06} Chen, C.~H., et al.\ 2006, 
\apjs, 166, 351
\bibitem[Chiar et al.(2007)]{chiar07} Chiar, J.~E., et al.\ 
2007, \apjl, 666, L73
\bibitem[Chihara et al.(2002)]{chi02} Chihara, H., Koike, C., 
         Tsuchiyama, A., Tachibana, S., \& Sakamoto, D.\ 2002, 
         \aap, 391, 267
\bibitem[Cohen et al.(2003)]{coh03} Cohen, M., Megeath, 
S.~T., Hammersley, P.~L., Mart{\'{\i}}n-Luis, F., 
\& Stauffer, J.\ 2003, \aj, 125, 2645
\bibitem[Correia et 
al.(2006)]{corr06} Correia, S., Zinnecker, H., Ratzka, T., \& Sterzik, M.~F.\ 2006,
\aap, 459, 909
\bibitem[D'Alessio et al.(2001)]{daless01} D'Alessio, P., 
        Calvet, N., \& Hartmann, L.\ 2001, \apj, 553, 321
\bibitem[D'Alessio et al.(2005)]{daless05} D'Alessio, P., 
         et al.\ 2005, \apj, 621, 461
\bibitem[de Pater 
\& Lissauer(2001)]{depater01} de Pater, I., \& Lissauer, J.~J.\ 2001, Planetary
Sciences, by Imke de Pater and Jack J.~Lissauer, pp.~544.~ISBN
0521482194.~Cambridge, UK: Cambridge University Press, December 2001.,
\bibitem[de Zeeuw et al.(1999)]{dezee99} de Zeeuw, P.~T., 
Hoogerwerf, R., de Bruijne, J.~H.~J., Brown, A.~G.~A., 
\& Blaauw, A.\ 1999, \aj, 117, 354
\bibitem[Desch \& Connolly(2002)]{dc02} Desch, S.~J., \& Connolly, H.~C.\ 
         2002, Meteoritics \& Planet. Sci., 37, 183
\bibitem[Dodd(1981)]{dodd81} Dodd, R.~T.\ 1981, {\it Meteorites - A 
petrologic-chemical synthesis}, .~Cambridge University Press, 1981.~368
\bibitem[Dorschner et al.(1995)]{dor95} Dorschner, J., 
         Begemann, B., Henning, T., Jaeger, C., 
         \& Mutschke, H.\ 1995, \aap, 300, 503
\bibitem[Draine(2003)]{d03} Draine, B.~T.\ 2003, \araa, 41, 241
\bibitem[Edgar et al.(2008)]{edgar08} Edgar, R.~G., Nordhaus, 
J., Blackman, E.~G., \& Frank, A.\ 2008, \apjl, 675, L101
\bibitem[Emery et al.(2007)]{emery07} Emery, J.~P., Dalle Ore, 
C.~M., Cruikshank, D.~P., Fern{\'a}ndez, Y.~R., Trilling, D.~E., 
\& Stansberry, J.~A.\ 2007, Bulletin of the American Astronomical Society, 38, 510
\bibitem[Fabian et al.(2000)]{fab00} Fabian, D., J{\"a}ger, C., 
         Henning, T., Dorschner, J., 
         \& Mutschke, H.\ 2000, \aap, 364, 282
\bibitem[Fabian et al.(2001)]{fab01} Fabian, D., Henning, T., 
         J{\" a}ger, C., Mutschke, H., Dorschner, J., 
         \& Wehrhan, O.\ 2001, \aap, 378, 228
\bibitem[Furlan et al.(2006)]{fur06} Furlan, E., et al.\ 
2006, \apjs, 165, 568 
\bibitem[Gail(2004)]{gail04} Gail, H.-P.\ 2004, \aap, 413, 571
\bibitem[Gatti et 
al.(2006)]{gatti06} Gatti, T., Testi, L., Natta, A., Randich, S., \& Muzerolle, J.\
2006, \aap, 460, 547
\bibitem[Gervais \& Piriou(1975)]{gp75} Gervais, F., \& 
Piriou, B.\ 1975, \prb, 11, 3944 
\bibitem[Ghez et al.(1993)]{ghez93} Ghez, A.~M., Neugebauer, 
G., \& Matthews, K.\ 1993, \aj, 106, 2005
\bibitem[Gr{\"u}n et al.(2001)]{grun01} Gr{\"u}n, E., 
Gustafson, B.~A.~S., Dermott, S., 
\& Fechtig, H.\ 2001, Interplanetary Dust, Edited by E.~Gr{\"u}n, 
B.A.S.~Gustafson, S.~Dermott, and H.~Fechtig.~Astronomy and Astrophysics 
Library.~2001, 804 p., ISBN: 3-540-42067-3.~ Berlin: Springer, 2001.,
\bibitem[Guan et al.(2000)]{guan00} Guan, Y., Huss, G.~R., 
MacPherson, G.~J., \& Wasserburg, G.~J.\ 2000, Science, 289, 1330
\bibitem[Guenther et 
al.(2007)]{guen07} Guenther, E.~W., Esposito, M., Mundt, R., Covino, E., Alcal{\'a},
J.~M., Cusano, F., \& Stecklum, B.\ 2007, \aap, 467, 1147
\bibitem[Hallenbeck et al.(1998)]{hnd98} Hallenbeck, S.~L., 
Nuth, J.~A., \& Daukantas, P.~L.\ 1998, Icarus, 131, 198
\bibitem[Hallenbeck \& Nuth(1998)]{hn98} Hallenbeck, S., \& Nuth, 
J.\ 1998, \apss, 255, 427
\bibitem[Hanner(2003)]{hanner03} Hanner, M.~S.\ 2003, 
Astromineralogy, 609, 171
\bibitem[Hanner \& Bradley(2004)]{habr04} Hanner, M.~S., \& Bradley, 
J.~P.\ 2004, Comets II, 555
\bibitem[Harker \& Desch(2002)]{hd02} Harker, D.~E., \& Desch, S.~J.\ 
         2002, \apjl, 565, L109
\bibitem[Harker et al.(2002)]{har02} Harker, D.~E., Wooden, D.~H., 
         Woodward, C.~E., \& Lisse, C.~M.\ 2002, \apj, 580, 579
\bibitem[Heaney et al.(1994)]{hean94} Heaney, P.~J.\ (1994) Structure 
and Chemistry of the Low-Pressure Silica Polymorphs. In {\it Silica: 
Physical Behavior, Geochemistry and Materials Applications}, Reviews in 
Mineralogy (eds. P.~J.\ Heaney, C.~T.\ Prewitt, \& G.~V. Gibbs). 
Mineralogical Society of America, Washington, D.~C., vol. 29, chap. 1, 
pp 1-40
\bibitem[Heaton(1971)]{hea71} Heaton, H.~I.\ 1971, Journal of the Optical
Society of America, 61, 275
\bibitem[Hemley et al.(1994)]{hemley94} Hemley, R.~J., Prewitt, C.~T., 
\& Kingma, K.~J.\ (1994) High-Pressure Behavior of Silica. In {\it Silica: 
Physical Behavior, Geochemistry and Materials Applications}, Reviews in 
Mineralogy (eds. P.~J.\ Heaney, C.~T.\ Prewitt, \& G.~V. Gibbs). 
Mineralogical Society of America, Washington, D.~C., vol. 29, chap. 2, 
pp 41-81
\bibitem[Henning \& Mutschke(1997)]{hm97} Henning, T., \& Mutschke, H.\ 
1997, \aap, 327, 743
\bibitem[Henning et al.(1999)]{hen99} Henning, T., Il'In, V.~B., 
         Krivova, N.~A., Michel, B., \& Voshchinnikov, N.~V.\ 1999, 
         \aaps, 136, 405
\bibitem[Herbig 
\& Bell(1988)]{hbc88} Herbig, G.~H., \& Bell, K.~R.\ 1988, Lick Observatory
Bulletin, Santa Cruz: Lick Observatory, |c1988,
\bibitem[Higdon et al.(2004)]{hig04} Higdon, S.~J.~U., et al.\ 2004, 
         \pasp, 116, 975
\bibitem[Hofmeister et al.(1992)]{hof92} Hofmeister, A.~M., Rose, T.~P.,
         Hoering, T.~C., \& Kushiro, I.\ 1992, Journal of Physical Chemistry , 96,
10213
\bibitem[Honda et al.(2003)]{honda03} Honda, M., Kataza, H., 
         Okamoto, Y.~K., Miyata, T., Yamashita, T., Sako, S., 
         Takubo, S., \& Onaka, T.\ 2003, \apjl, 585, L59
\bibitem[Houck et al.(2004)]{hou04} Houck, J.~R., et al.\ 
         2004, \apjs, 154, 18
\bibitem[Jaeger et al.(1998)]{jag98} Jaeger, C., Molster, F.~J., 
Dorschner, J., Henning, T., Mutschke, H., \& Waters, L.~B.~F.~M.\ 
1998, \aap, 339, 904
\bibitem[Jensen et al.(1996)]{jensen96} Jensen, E.~L.~N., 
Mathieu, R.~D., \& Fuller, G.~A.\ 1996, \apj, 458, 312
\bibitem[Jensen 
\& Mathieu(1997)]{jm97} Jensen, E.~L.~N., \& Mathieu, R.~D.\ 1997, \aj, 114, 301
\bibitem[Kamp 
\& Dullemond(2004)]{kadull04} Kamp, I., \& Dullemond, C.~P.\ 2004, \apj, 615, 991
\bibitem[Kastner et al.(2006)]{kast06} Kastner, J.~H., 
Buchanan, C.~L., Sargent, B., \& Forrest, W.~J.\ 2006, \apjl, 638, L29
\bibitem[Kemper et al.(2004)]{kemp04} Kemper, F., Vriend, W.~J., 
        \& Tielens, A.~G.~G.~M.\ 2004, \apj, 609, 826 (erratum: 633, 534)
\bibitem[Kenyon et al.(1994)]{ken94} Kenyon, S.~J., 
Dobrzycka, D., \& Hartmann, L.\ 1994, \aj, 108, 1872
\bibitem[Kenyon \& Hartmann(1995)]{kh95} Kenyon, S.~J., \& 
         Hartmann, L.\ 1995, \apjs, 101, 117
\bibitem[Kimura \& Nuth(2007)]{kn07} Kimura, Y., \& Nuth, J.~A., 
III 2007, \apj, 664, 1253
\bibitem[Klein \& Hurlbut(1977)]{klhurl97} Klein, C., \& Hurlbut, C.~S.,\ Jr.\ 1977 
{\it Manual of Mineralogy}, John Wiley and Sons, Inc., 21, 681
\bibitem[Koike \& Hasegawa(1987)]{koike87} Koike, C., \& Hasegawa, 
H.\ 1987, \apss, 134, 361
\bibitem[Koike et al.(1989)]{koike89} Koike, C., Komatuzaki, 
T., Hasegawa, H., \& Asada, N.\ 1989, \mnras, 239, 127
\bibitem[Li \& Draine(2001)]{ld01} Li, A., \& Draine, B.~T.\ 
         2001, \apjl, 550, L213
\bibitem[Li \& Draine(2002)]{ld02} Li, A., \& Draine, B.~T.\ 2002, \apj, 564, 803
\bibitem[Lisse et al.(2007)]{lisse07} Lisse, C.~M., Kraemer, 
K.~E., Nuth, J.~A., Li, A., \& Joswiak, D.\ 2007, Icarus, 191, 223
\bibitem[Luhman(2004)]{luh04} Luhman, K.~L.\ 2004, \apj, 602, 816
\bibitem[Luhman(2008)]{luh08} Luhman, K.~L.\ 2008, in ASP Conf. Ser. : 
Handbook of Star Forming Regions, submitted
\bibitem[MacKinnon \& Rietmeijer(1987)]{macriet87} MacKinnon, I.~D.~R., \& 
Rietmeijer, F.~J.~M.\ 1987, Reviews of Geophysics, 25, 1527
\bibitem[Marshall et al.(2007)]{marsh07} Marshall, J.~A., 
Herter, T.~L., Armus, L., Charmandaris, V., Spoon, H.~W.~W., Bernard-Salas, 
J., \& Houck, J.~R.\ 2007, \apj, 670, 129
\bibitem[Mathieu et al.(1989)]{math89} Mathieu, R.~D., Walter, 
F.~M., \& Myers, P.~C.\ 1989, \aj, 98, 987
\bibitem[McClure et al.(2008)]{mcc08} McClure, M.~K., et al.\ 
        2008, in preparation
\bibitem[Merrill(1974)]{merr74} Merrill, K.~M.\ 1974, Icarus, 
23, 566
\bibitem[Metchev(2006)]{metch06} Metchev, S.~A.\ 2006, 
Ph.D.~Thesis,
\bibitem[Mikouchi et al.(2007)]{mikou07} Mikouchi, T., 
Tachikawa, O., Hagiya, K., Ohsumi, K., Suzuki, Y., Uesugi, K., Takeuchi, A., 
\& Zolensky, M.~E.\ 2007, Lunar and Planetary Institute Conference Abstracts, 38, 1946
\bibitem[Montmerle et al.(1983)]{mont83} Montmerle, T., 
Koch-Miramond, L., Falgarone, E., \& Grindlay, J.~E.\ 1983, \apj, 269, 182
\bibitem[Mutschke et al.(1998)]{mut98} Mutschke, H., 
Begemann, B., Dorschner, J., Guertler, J., Gustafson, B., Henning, 
T., \& Stognienko, R.\ 1998, \aap, 333, 188
\bibitem[Norton(2002)]{nort02} Norton, O.~R.\ 2002, The 
Cambridge Encyclopedia of Meteorites, by O.~Richard Norton, pp.~374.~ISBN 
0521621437.~Cambridge, UK: Cambridge University Press, March 2002.,
\bibitem[Palik(1985)]{palik85} Palik, E.~D.\ 1985, Academic 
Press Handbook Series, New York: Academic Press, 1985, edited by Palik, 
Edward D.,
\bibitem[Pascucci et al.(2007)]{pas07} Pascucci, I., et al.\ 
2007, \apj, 663, 383
\bibitem[Pilipp et al.(1998)]{pil98} Pilipp, W., Hartquist, T.~W., 
         Morfill, G.~E., \& Levy, E.~H.\ 1998, \aap, 331, 121
\bibitem[Plyusnina et al.(1970)]{ply70} Plyusnina, I.~I., Maleyev, N.~M. 
\& Yefimova, G.~A.\ 1970, J Solid State Chem, 44, 24
\bibitem[Posch et al.(2007)]{posch07} Posch, T., Mutschke, H., 
Trieloff, M., \& Henning, T.\ 2007, \apj, 656, 615
\bibitem[Prato 
\& Simon(1997)]{ps97} Prato, L., \& Simon, M.\ 1997, \apj, 474, 455
\bibitem[Preibisch et 
al.(1998)]{prei98} Preibisch, T., Guenther, E., Zinnecker, H., Sterzik, M., Frink,
S., \& Roeser, S.\ 1998, \aap, 333, 619
\bibitem[Ratzka et 
al.(2005)]{ratz05} Ratzka, T., K{\"o}hler, R., \& Leinert, C.\ 2005, \aap, 437, 611
\bibitem[Reipurth 
\& Zinnecker(1993)]{rz93} Reipurth, B., \& Zinnecker, H.\ 1993, \aap, 278, 81
\bibitem[Rietmeijer \& McKay(1985)]{rimc85} Rietmeijer, F.~J.~M., \& McKay, D.~S.\ 
1985, Meteoritics, 20, 743
\bibitem[Rietmeijer \& McKay(1986)]{rimc86} Rietmeijer, F.~J.~M., \& McKay, D.~S.\ 
1986, Lunar and Planetary Institute Conference Abstracts, 17, 710
\bibitem[Rietmeijer et al.(1986)]{riet86} Rietmeijer, 
F.~J.~M., Nuth, J.~A., \& MacKinnon, I.~D.~R.\ 1986, Icarus, 66, 211
\bibitem[Rietmeijer et al.(1999)]{riet99} Rietmeijer, 
F.~J.~M., Nuth, J.~A., III, \& Karner, J.~M.\ 1999, \apj, 527, 395
\bibitem[Rietmeijer et al.(2002)]{riet02} Rietmeijer, F.~J.~M., Hallenbeck, 
S.~L., Nuth, J.~A., \& Karner, J.~M.\ 2002, Icarus, 156, 269
\bibitem[Rokita et al.(1998)]{rok98} Rokita, M., Handke, M., \& Mozgawa, 
         W.\ 1998, Journal of Molecular Structure , 450, 213
\bibitem[Sargent et al.(2006)]{sarg06} Sargent, B., et al.\ 
2006, \apj, 645, 395 
\bibitem[Sargent et al.(2008)]{sarg08} Sargent, B., et al.\ 
        2008, in preparation
\bibitem[Schaefer et al.(2006)]{schaef06} Schaefer, G.~H., 
Simon, M., Beck, T.~L., Nelan, E., \& Prato, L.\ 2006, \aj, 132, 2618
\bibitem[Servoin \& Piriou(1973)]{sp73} Servoin, J.~L., \& Piriou, B.\ 1973, 
Physica Status Solidi B Basic Research , 55, 677
\bibitem[Shu et al.(1996)]{ssl96} Shu, F.~H., Shang, H., 
         \& Lee, T.\ 1996, Science, 271, 1545
\bibitem[Silverstone et al.(2006)]{sil06} Silverstone, M.~D., 
et al.\ 2006, \apj, 639, 1138
\bibitem[Simon et al.(1996)]{sim96} Simon, M., Holfeltz, 
S.~T., \& Taff, L.~G.\ 1996, \apj, 469, 890
\bibitem[Sogawa et al.(2006)]{sog06} Sogawa, H., Koike, C., 
Chihara, H., Suto, H., Tachibana, S., Tsuchiyama, A., \& Kozasa, T.\ 2006, 
\aap, 451, 357
\bibitem[Spitzer \& Kleinman(1961)]{sk61} Spitzer, W.~G., \& 
Kleinman, D.~A.\ 1961, Physical Review , 121, 1324
\bibitem[Steyer et al.(1974)]{steyer74} Steyer, T.~R., Day, 
K.~L., \& Huffman, D.~R.\ 1974, \ao, 13, 1586
\bibitem[Swainson 
\& Dove(1993)]{swdo93} Swainson, I.~P., \& Dove, M.~T.\ 1993, Physical Review
Letters, 71, 193
\bibitem[Takami et 
al.(2003)]{tak03} Takami, M., Bailey, J., \& Chrysostomou, A.\ 2003, \aap, 397, 675
\bibitem[Tuttle \& Bowen(1958)]{tb58} Tuttle, O.~F., \& Bowen, N.~L.\ 1958, 
Geol.\ Soc.\ Am.\ Mem.\ , 74, 1-153
\bibitem[Uchida et al.(2004)]{uch04} Uchida, K.~I., et al.\ 
         2004, \apjs, 154, 439
\bibitem[Walter(1992)]{walt92} Walter, F.~M.\ 1992, \aj, 104, 
758
\bibitem[Watson et al.(2007)]{wat07} Watson, D.~M., et al.\ 
2007, \nat, 448, 1026
\bibitem[Wenrich \& Christensen(1996)]{wc96} Wenrich, M.~L., 
         \& Christensen, P.~R.\ 1996, \jgr, 101, 15921
\bibitem[Werner et al.(2004)]{wer04} Werner, M.~W., et al.\ 
         2004, \apjs, 154, 1
\bibitem[Wilking et al.(2005)]{wilk05} Wilking, B.~A., Meyer, 
M.~R., Robinson, J.~G., \& Greene, T.~P.\ 2005, \aj, 130, 1733
\bibitem[Williams et al.(1993)]{wi93} Williams, Q., Hemley, 
R.~J., Kruger, M.~B., \& Jeanloz, R.\ 1993, \jgr, 98, 22157
\bibitem[Zolensky et al.(2006)]{zol06} Zolensky, M.~E., et 
al.\ 2006, Science, 314, 1735

\end{thebibliography}
\end{document}